\documentclass[12pt]{article}
\pdfoutput=1
\usepackage[dvipsnames,table,xcdraw]{xcolor}
\usepackage{draft} 
\usepackage{hyperref}
\usepackage{url}
\usepackage{graphicx,color,subfig,upgreek,mathtools}
\usepackage{amsfonts,amssymb,amsmath,multirow}
\usepackage{mathabx,epsfig}
\usepackage{pifont}
\usepackage{mciteplus}
\usepackage{skak}
\usepackage{bbm}
\usepackage{bm}
\DeclareFontFamily{OT1}{pzc}{}
\DeclareFontShape{OT1}{pzc}{m}{it}{<-> s * [1.10] pzcmi7t}{}
\DeclareMathAlphabet{\mathpzc}{OT1}{pzc}{m}{it}

\def\be#1\ee{\begin{align}#1\end{align}}

\usepackage{tikz-cd}
\usepackage{datetime}

\def\tilde{\widetilde}

\def\^{{\wedge}}
\def\*{{\star}}
\renewcommand\maketitle{}
\definecolor{ao}{rgb}{0.13, 0.55, 0.13}

\begin{document}

\begin{titlepage}

\begin{center}

\hfill \\
\hfill \\
\vskip 0.5cm

\title{\vspace{-1cm} 
Anomalies of Average Symmetries:  \\Entanglement and Open Quantum Systems
}

\author{Po-Shen Hsin$^{1}$, Zhu-Xi Luo$^{2}$, Hao-Yu Sun$^{3}$}

\address{${}^1$Mani L. Bhaumik Institute for Theoretical Physics,\\
Department of Physics and Astronomy,\\
University of California Los Angeles, Los Angeles, CA 90095, USA}

\address{${}^2$Department of Physics, Harvard University, Cambridge, MA 02138, USA}

\address{${}^3$ 
Weinberg Institute, Department of Physics,\\
The University of Texas at Austin, Austin, TX 78712-1192, USA
}

\end{center}

\abstract{
Symmetries and their anomalies are powerful tools for understanding quantum systems. However, realistic systems are often subject to disorders, dissipation and decoherence. In many circumstances, symmetries are not exact but only on average.  This work investigates the constraints on mixed states resulting from non-commuting average symmetries. We will focus on the cases where the commutation relations of the average symmetry generators are violated by nontrivial phases, and call such average symmetry anomalous. 
We show that anomalous average symmetry implies degeneracy in the density matrix eigenvalues, and present several lattice examples with average symmetries including XY chain, Heisenberg chain, and deformed toric code models. In certain cases, the results can be further extended to reduced density matrices, leading to a new lower bound on the entanglement entropy. We discuss several applications in the contexts of many body localization, quantum channels, entanglement phase transitions and also derive new constraints on the Lindbladian evolution of open quantum systems.  
}

\vfill

\today

\vfill

\end{titlepage}


\tableofcontents

\unitlength = .8mm

\setcounter{tocdepth}{3}

\medskip

\section{Introduction}
\label{sec:intro}

Symmetry has established itself as a crucial tool for understanding quantum systems. In particular, anomalies of symmetries provide powerful constraints on the dynamics,  such as interactions in low energy effective field theories ({\it e.g.} \cite{Witten:1983tw}),
Lieb--Schultz--Mattis type theorems \cite{LIEB1961407,PhysRevLett.84.3370,Hastings_2004,Else:2019lft,Cheng:2022sgb}, deconfinement in gauge theories ({\it e.g.} \cite{Shimizu:2017asf,Hsin:2019fhf,Komargodski:2020mxz}), and the nontrivial boundary states of symmetry-protected topological phases (SPTs) such as topological insulators and topological superconductors ({\it e.g.} \cite{PhysRevB.87.155114,Else:2014vma,Witten_2016}).

The global symmetry $G$ in quantum systems can be realized in different ways on the Hilbert space and in the presence of defects that implement symmetry transformations. In such cases, the symmetry is anomalous, and
the anomalies of symmetries can be represented by nontrivial commutation relations of the symmetry generators violating the group relations in $G$ by nontrivial phases (see {\it e.g.} \cite{Barkeshli:2022edm,Barkeshli:2023bta} and references therein). For instance, suppose symmetry $G=\mathbb{Z}_2\times \mathbb{Z}_2$, which commutes with the Hamiltonian, realizes projectively in a quantum mechanical system as the Pauli $X$ and Pauli $Z$ operators satisfying $XZ=-ZX.$ The symmetry generators acting on the Hilbert space describes an extension $\tilde G$ of $G$ by extra phases in the commutation relations. In this example, the phase is simply a sign, and $\tilde G=\mathbb{D}_8$, the Dihedral group of order 8.
The commutation relation implies that 
a symmetry-invariant subspace cannot be one-dimensional. For example, suppose that there is an energy gap separating the ground state subspace and excited states. If
there was a unique ground state $|0\rangle$, the state $Z|0\rangle$ would be a different ground state since it has the same energy as $|0\rangle$ does but has a different eigenvalue under $X$. Thus the commutation relation of the symmetry generators implies the degeneracy of ground states.

Real quantum systems in experiments are often affected by decoherence, disorders and dissipation. Instead of pure states, it is thus natural to consider mixed states and the presence of ambient environments. In such cases, symmetries may not be present exactly, but only on average. In this work, we will be interested in average symmetries on mixed state density matrices, instead of on Hamiltonians or Lagrangians which are not suitable for describing generic out-of-equilibrium systems. 
A density matrix $\rho$ is said to have a unitary average symmetry $U$ if it satisfies
\begin{equation}
\label{eq:def}
    U\rho U^{-1}=\rho~.
\end{equation}
Such symmetries are also referred to as weak symmetries (see {\it e.g.} \cite{PhysRevA.98.042118}). 
The above average symmetry condition is defined to preserve the expectation value of every operator ${\cal O}$ with respect to the state $\rho$:
\begin{equation}
    \text{Tr}\left(\rho {\cal O}\right)=    \text{Tr}\left(\rho {\cal O}'\right),\quad {\cal O}'=U^{-1}{\cal O} U~.
\end{equation} 
For instance, if the density matrix $\rho$ describes the thermal equilibrium at finite temperature $T=1/\beta$, {\it i.e.} $\rho=\frac{1}{Z_\beta}e^{-\beta H}$ for partition function $Z_\beta$, the symmetry of the Hamiltonian $UH=HU$ would then be an average symmetry of the thermal equilibrium density matrix. We will discuss more general density matrices that are not necessarily related to thermal equilibrium.

Since the generators of average symmetry act on the density matrix by conjugation (\ref{eq:def}), one cannot directly apply the reasoning that leads to degeneracy for pure states as we reviewed in the beginning. For example, while the Pauli $X$ and $Z$ operators do not commute, conjugating the density operator $\rho$ by $X$ commutes with conjugating $\rho$ by $Z$. Naively, it might seem that all the information about phases in the commutation relations of average symmetry generators is lost, as conjugating the density operator by phases always recover the density matrix itself. On the contrary, we will show that the nontrivial commutation relations can still lead to important physical constraints on the mixed state systems.

We remark that
there are various recent progress to investigate the consequences of such symmetries on mixed states.
Examples of such systems include average symmetry-protected topological (SPT) phases \cite{Ma:2022pvq,Zhang:2022jul,ma2023topological,Lee:2022hog},  mixed state topological orders \cite{PhysRevB.95.075106}, and ensemble averages in Sachdev--Ye--Kitaev (SYK) models \cite{Sachdev:2015efa,kitaevTalks,kitaevkitp1,kitaevkitp,Maldacena:2016hyu,Kitaev:2017awl,Qi:2021oni,Antinucci:2023uzq}. 
In particular, \cite{Ma:2022pvq,Zhang:2022jul,ma2023topological} discussed examples with mixed anomalies between exact symmetries and average symmetries. In \cite{Kawabata:2023dlv}, examples of Lieb--Schultz--Mattis type theorems are discovered in open quantum systems with lattice translation and exact $U(1)$ symmetry under Lindbladian evolution. However, a complete understanding of the implications for average symmetry on mixed states is still lacking.

When the density operator is proportional to the identity matrix $\rho=\frac{1}{N}\mathbf{1}$ for Hilbert space of total dimension $N$, {\it i.e.} when all eigenvalues of the density operator are exactly equal, the constraint from symmetry (\ref{eq:def}) is automatically satisfied.
In general, average symmetry will constrain how the density operator is ``close to being identity'', {\it i.e.} constrain the degeneracy of the eigenvalues of the density operator. In other words, it is a constraint on the purity of the density matrix. 
Such a degeneracy has applications to new bounds on entanglement entropy, as we will discuss later. 

In this work we will develop a general method and derive new constraints on the systems using average symmetry. 
We will focus on lattice systems where the total Hilbert space is the tensor product of finite-dimensional local Hilbert spaces. 
The input data of our discussion is the following:
\begin{itemize}
    \item[1. ] Average symmetry $G$ of a system of mixed states. The symmetry $G$ is described by symmetry generators that act non-trivially on the Hilbert space, {\it i.e.} they are not complex numbers but nontrivial operators.
    \item[2. ]  
    The commutation relations of the symmetry generators that are violated by $U(1)$ phases, thereby forming an extension $\tilde G$. 
\end{itemize}
We will call average symmetries whose generators act on the system with nontrivial phases that violate their commutation relations ``anomalous'' average symmetries, similar to exact symmetries with nontrivial 't Hooft anomalies. 
We note that the 't Hooft anomalies for unitary symmetries can be detected by the nontrivial phases in the commutation relations between the symmetry generators in the presence of symmetry fluxes (see {\it e.g.} \cite{Cheng:2022sgb,Seiberg:2023cdc}).
The anomaly of average symmetry can be characterized by the dimension $d_M$ of the minimal projective representation of $G$ that manifests the nontrivial phases in the commutation relations of the symmetry generators.
When the average symmetry is anomalous, {\it i.e.} the phases are nontrivial, the dimension must be greater than one:
\begin{equation}
    \text{Anomalous average symmetry}:\quad d_M>1~.
\end{equation}
In particular, the symmetry cannot act trivially while respecting the anomalous commutation relations.

The average symmetry condition \eqref{eq:def} implies that the density matrix is in the center of $\tilde{G}$. By Schur's lemma \cite{Schur}, an anomalous average symmetry implies 
nontrivial degeneracy of the density matrix eigenvalues.
As a simple example, consider the quantum mechanics of a single spin-$1/2$ magnet. Suppose the density matrix, a two-by-two matrix, has an average symmetry generated by Pauli matrices $X$ and $Z$, then the density matrix will be a maximally mixed state: the eigenvalues have degeneracy $d_M=2$. 
In Section \ref{sec:densitymatrix}, , we discuss in detail the consequences of anomalous average symmetry on density matrices.

We remark that the constraints we derived from the anomalous average symmetry do not require the presence of additional exact (or strong) symmetries. 
This is to be contrasted with the approaches in \cite{Ma:2022pvq,Zhang:2022jul,ma2023topological} which examine the possibilities of average SPT phases when there is some exact symmetry which have mixed anomalies with average symmetry on the boundary of bulk with nontrivial topological properties.\footnote{
In our discussion we do not construct SPT phases with average symmetry. It would be interesting to investigate the bulk-boundary correspondence for the boundary systems with anomalous average symmetry.
}

We will also discuss other applications including many body localization, quantum channels, as well as the Lindbladian evolution.

\subsection{Summary of results}

Below we outline some of the results and examples for readers' convenience. 

\subsubsection{New bound on entanglement entropy}

In Section \ref{sec:entanglement}, 
we apply our method to derive new bounds on the entanglement entropy for density operators with anomalous average symmetry. We show that the presence of anomalous average symmetry gives a lower bound of the von Neumann entropy:
\begin{equation}
    S(\rho)\geq \log d_M~,
\end{equation}
where $d_M>1$ is the dimension of the minimal projective representation of anomalous average symmetry $G$ with commutation relation between symmetry generators violated by phases as described by the extension $\tilde G$. In particular, density matrices with anomalous average symmetry cannot represent pure states, since the density matrices for pure states would have zero von Neumann entropy.
Similarly, we show the $\alpha$-R{\'e}nyi entropy $S_\alpha(\rho)$ for all real $\alpha>1$ is also bounded from below by $\log d_M$ independent of $\alpha$.

As an application, we can bound the entanglement entropy of a subregion $A$. We start with a density matrix $\rho$ of the entire system with average symmetry (for instance, the density matrix can be a pure state, and the average symmetry comes from an exact symmetry of the pure state). Then the reduced density matrix $\rho_A$ on region $A$ will also have average symmetry: for average symmetry generator $U$ factorizable as $U=U_A\otimes U_B$ for $U_A$ supported only on region $A$ and $U_B$ supported on the complement region, such as the case of on-site anomalous average symmetries, the reduced density matrix has average symmetry $U_A$.\footnote{
There are cases where the factorization property does not apply, such as quenched disorders, then the reduced density matrix does not have average symmetry and the entanglement entropy cannot be constrained in this way. This is discussed in section \ref{sec:ensemble}.
} We note that the average symmetry $U_A$ acting on subregion can be anomalous even if the average symmetry $U$ is non-anomalous, as we will show in examples in Section \ref{sec:entanglement}.
Let us denote the dimension of the minimal projective representation realizing the anomalous commutation relation by $d_M^A$. Then the entanglement entropy of region $A$ satisfies
\begin{equation}
    S(\rho_A)\geq \log d_M^A~.
\end{equation}
Similarly, the $\alpha$-R{\'e}nyi entropy $S_\alpha(\rho_A)$ for all real $\alpha>1$ is also bounded from below by $\log d_M^A$.
We can rephrase the bound on entanglement entropy as the following statement:
\begin{equation}
\fbox{
\text{Anomalous Average Symmetries are source of Entanglement Entropy.}
}~
\end{equation}
The anomalous average symmetry can depend on the region geometry, and thus the bound constrains the scaling of the entanglement entropy with the region size. 

We give various examples illustrating the bound. 
For instance, we show that in any (2+1)d system with a density matrix that has anomalous average $\mathbb{Z}_2$ one-form symmetries generated by ``electric'' and ``magentic'' loop operators of order 2 with mutual $\pi$ statistics, the entanglement entropy of a region $A$ with boundary size $L$ (measured in the unit of lattice spacing 
at the scale where these average symmetries emerge
) satisfies
\begin{equation}
    S(\rho_A)\geq L\log 2-\log 2~.
\end{equation}
We remark that the bound is saturated for the exactly solvable toric code model ground state, with $\log 2$ equal to the topological entanglement entropy \cite{Levin_2006,Kitaev_2006} of the $\mathbb{Z}_2$ toric code topological order. The result depends only on the density matrix and its average symmetry, while it is independent of the choice of Hamiltonian.

Another example is a family of deformed $\mathbb{Z}_2$ toric code model with $\mathbb{Z}_2$ one-form symmetries generated by the electric and magnetic line operators. The family is labelled by a sublattice where the electric charge is condensed on the sites of the sublattice.
The system can be further deformed to be not exactly solvable as long as the one-form symmetries are preserved. We show that by adjusting the ratio of the sites where the electric charge is condensed, the ground state entanglement entropy  undergoes a transition from area law to volume law.

\subsubsection{Applications to many-body localization}

When the number of symmetry generators grow with the system size, {\it i.e.} there are local conserved quantities, the dimension of the minimal projective representation $d_M$ for the reduced density matrix can also grow with the system size. 
 As an application, we show that the density matrix with local average symmetry that is non-Abelian, {\it i.e.} anomalous, must have volume law entanglement entropy. This is consistent with the non-Abelian symmetry obstruction to many body localization \cite{PhysRevB.96.041122,PhysRevB.94.224206}.

We remark that the phenomenon is similar to the ``anomalous'' system protected from many body localization on the boundary of ``statistical topological insulators" \cite{PhysRevB.89.155424,PhysRevB.92.085139}. An example is the weak (3+1)d topological insulator from stacking (2+1)d topological insulators, with time-reversal symmetry as well as an average translation symmetry along the stacking direction due to disorder averaging.
The boundary of such (3+1)d disordered weak topological insulator is protected against Anderson localization even in the presence of strong disorder due to the mixed anomaly between
time-reversal symmetry and average translation symmetry \cite{PhysRevB.86.045102,PhysRevLett.108.076804}.

\subsubsection{Constraints on quantum channels}

In Section \ref{sec:quantumchannel},
we show that symmetry can constrain the quantum channel in the Kraus representation \cite{kraus1983states,nielsen2010quantum}
\begin{equation}
    {\cal E}[\rho]=\sum_i K_i \rho K_i^\dag~,
\end{equation}
where $K_i$ are the Kraus operators that satisfy $\sum_i K_iK_i^\dag=1$. 
One can define two kinds of average symmetries of quantum channel: 
\begin{itemize}
    \item [(1)]
The transformed density matrix ${\cal E}[\rho]$ has an average symmetry for any initial density matrix $\rho$. 
The symmetry acts on the Kraus operators as a unitary representation
\begin{equation}
    UK_j=\sum_{l} u_{jl}K_l~,
\end{equation}
where $(u_{ij})$ is a constant unitary matrix multiplied by the identity operator. 

\item[(2)] The quantum channel preserves the average symmetry of the initial density matrix $U\rho U^{-1}=\rho~,$ 
and the Kraus operator satisfy
$
    UK_iU^{-1}=\sum_{l} u_{jl}K_l~,
$
where $(u_{ij})$ is a constant unitary matrix multiplied the identity operator.
The density matrix ${\cal E}[\rho]$ still has average symmetry $U$ for the density matrix $\rho$.
\end{itemize}
The second case has been considered in the literature such as in ref. \cite{de_Groot_2022}. We will focus on the first case here: When the symmetry $U$ is anomalous with dimension $d_M$ for the minimal projective representation, the quantum channels with symmetry in the first case have at least $d_M$ Kraus operators.
We can use average symmetry to constrain the quantum channel, leading to new lower bound on the von Neumann entropy and entanglement entropy 
of the evolved density matrix. In particular, this sheds light on the existence of entanglement phase transitions upon dialing the parameters in quantum channels where multiple drives of evolution are present, such as monitored circuits with competing measurements. We present several examples illustrating the idea. 
Our analysis further extends to the cases of approximate average symmetries in the channel, where the elementary quantum channel has no average symmetry, but an average symmetry emerges in the late-time limit after the elementary channel has been implemented for a large number of times.

\subsubsection{Constraints on open quantum system dynamics}

In Section \ref{sec:openquantumsystem}, we turn to the consequences of anomalous average symmetries on evolution in open quantum systems.
A large class of open quantum systems where the interactions with environments obey suitable assumptions such as being Markov processes can be described by the Lindbladian master equation 
$
    d\rho/dt={\cal L}[\rho]~
$ \cite{Lindblad:1975ef,Gorini:1975nb}, 
where ${\cal L}$ is the Lindbladian super-operator. Symmetries of the evolution can be described by the condition: 
\begin{equation}\label{eqn:Lindbladinsymmetry}
 U^{-1}\circ {\cal L}\circ U={\cal L}~,
\end{equation}
where $U$ in the above equation is regarded as a super-operator.
We can rewrite the symmetry condition on the Lindbladian ${\cal L}$ in terms of their actions on the density matrices:
\begin{equation}
    U^{-1}{\cal L}[U\rho]={\cal L}[\rho]~.
\end{equation}

Examples include SPT phases in open quantum systems \cite{de_Groot_2022}, and dissipative generalization \cite{Kawabata:2022cpr} of the SYK model.
Consequences of weak symmetry breaking in open quantum systems were investigated in \cite{PhysRevA.98.042118}, with applications to error correction in open quantum systems \cite{PhysRevLett.125.240405}. 

We show that anomalous symmetry for the Lindbladian evolution (\ref{eqn:Lindbladinsymmetry}) implies there must be at least $d_M$ number of blocks in the density matrix that grow or decay with the same rate. In particular, if the evolution has steady states, they must have degeneracy $\geq d_M$. Moreover, the evolution of a given density matrix must obey conservation law: decomposing the density matrix into different blocks under the symmetry transformation $\rho=\oplus_I\rho^{(I)}$, those blocks that grow or decay with the same nonzero rate must satisfy 
\begin{equation}
    \sum_I\text{Tr}\rho^{(I)}=0~,
\end{equation}
where the sum is over the blocks that grow or decay with the same nonzero rate. The condition ensures the trace of the density matrix is preserved under the evolution.

\subsection{Review of Schur's Lemma}

In the following analysis, we will repeatedly use Schur's lemma, which we rephrase in the following form for the reader's convenience:
\paragraph{Schur's Lemma} Consider a finite group $\tilde G$ and an $N$-dimensional decomposable representations $U$, {\it i.e.} 
there exists a similarity transformation $V$ such that $V^{-1} U V= U_1\oplus U_2$. If $U_1$  is irreducible and  has dimension $n$, then a diagonalizable $N$-by-$N$ matrix $M$ satisfying $MU(g)=U(g)M$ for any element $g\in \tilde G$ has $n$ degenerate eigenvalues. We will be particularly interested in the special case where $U$ is completely reducible, {\it i.e.} is decomposable into a direct sum of multiple irreducible representations $V^{-1} U V=\bigoplus_{i=1}^m U_i$ where $U_i$ is $n_i$-by-$n_i$. Then $MU=UM$ implies that $M$ has at most $m$ distinct eigenvalues, each with degeneracy $n_i$. As shown in \cite{CORNWELL199747}, all reducible representations of finite group, compact Lie group, or connected, non-compact, semi-simple Lie group are completely reducible. The same is true of any unitary reducible representation of any other group.
More details about Schur's lemma can be found in Appendix \ref{sec:Schur}.

\section{Consequences on Density Matrix}
\label{sec:densitymatrix}

In this section, we investigate the consequences of anomalous weak or average symmetries of a mixed-state density matrix $\rho$. As reviewed in the introduction, an average symmetry operator $U(g_a)$  is defined to act on $\rho$ by conjugation, 
$
U(g_a)\rho U(g_a)^{-1} = \rho. 
$
Here $g_a \in \tilde{G}$ is the abstract element of the symmetry group, while $U(g_a)$ is $g_a$ written in the representation $U$. 
When different symmetries $U(g_a)$ and $U(g_b)$ do not commute, the spectrum of $\rho$ can be degenerate. Below we will first describe the general recipe and then present examples which correspond host the usual symmetries of (1) the XY-spin chain, (2) the Heisenberg chain, as well as (3) the (2+1)d toric code. Our results hold for both conventional symmetries as well as higher symmetries. 

The general procedure consists of the following steps:
\begin{itemize}
\item Identify the symmetries and their commutation relations. For example, certain symmetry generators may not commute, giving rise to 
\be
U(g_a) U(g_b) U(g_a)^{-1} U(g_b)^{-1}= U(g_c),
\ee
where $U(g_c)$ is nontrivial. We will focus on the case where $\tilde{G}$ is a central extension of $G$, where  the extra element $g_c\in \tilde G$ is represented by a nontrivial phase, {\it i.e.} $U(g_c)=e^{\mathrm{i} \phi} \mathbbm{1}$ is a phase factor times the identity matrix. 
\item For all finite groups or compact Lie groups $\tilde{G}$, any representation $U(\tilde{G})$ is equivalent to the direct sum of irreducible representations via some similarity transformation $V$, {\it i.e.} $U'=V^{-1} U V =\bigoplus_{i=1}^m U_i$. Since similarity transformations preserve the trace as well as the identity matrix, the fact that $U(g_c)=e^{\mathrm{i} \phi} \mathbbm{1}$ means $U'(g_c)=e^{\mathrm{i} \phi} \mathbbm{1}$. 
\item Find out the irreducible representations $U_j$ of $\tilde{G}$ that satisfies the constraint $U_j(g_c)=e^{i\phi}\mathbbm{1}_{n_j}$, where $n_j\times n_j$ is the dimension of $U_j$. Such a representation cannot be one-dimensional because all numbers must commute with each other. 
\begin{itemize}
\item [-] If there is only one such irreducible representation, then Schur's lemma (reviewed in the introduction section and discussed in detail in Appendix \ref{sec:Schur}) indicates that the density matrix, which satisfies $U'(g) \rho = \rho U'(g)$, decomposes into blocks of $n_j\times n_j$ identity matrices, {\it i.e.} the full spectrum is $n_j$-fold degenerate. 
\item [-] If there are multiple irreducible representations that satisfy the constraint with different dimensions, then the spectrum can be $n_j$-fold degenerate wherever that multi-dimensional representation $U_j$ appears in the decomposition. In this case, one will need to work out the decomposition in order to see the degeneracy explicitly. 
\end{itemize}
\item After obtaining the information about the density matrix, one can use it to constrain other physical quantities such as entanglement entropy that we will discuss in section \ref{sec:entanglement}. 
\end{itemize}

We will discuss several examples below.

\subsection{Example: spin chains}

\subsubsection{Spin chains with symmetries of the XY model}
\label{sec:XY}

We start by considering a spin-1/2 chain of size $L$, where $L$ is odd.  Impose the following weak symmetries that are present in the XY-chain:
\be
U(g_z)=\bigotimes_i Z_i,\quad  U(g_x)=\bigotimes_i X_i. 
\ee
Here $X_i$, $Z_i$ are shorthand notations for Pauli matrices $\sigma^x_i, \sigma^z_i$ which act on the $i$-th qubit on the chain. We remark that the familiar term ``XY-chain" is used just for convenience and our analyses do not require the presence of a Hamiltonian. For a chain with an odd number of sites, these two symmetries anticommute and together generate the $D_8$ Dihedral group. 
The only irreducible representation of the $D_8$ group that satisfies
\be
U_2(g_z g_x g_z^{-1} g_x^{-1})= - \mathbbm{1}
\ee
is two-dimensional. Consequently, in a suitable basis, we must have
\be
U'(g)=\bigoplus_{i=1}^{2^{L-1}} U_i(g),
\ee
where each $U_i$ is the two-by-two dimensional irreducible representation of $D_8$. 
This case falls into the case (ii) in Appendix \ref{sec:Schur}. Consequently, the density matrix decomposes into a direct sum of two-by-two submatrices, each proportional to identity. The full spectrum has twofold degeneracy.

\subsubsection{Spin chains with symmetries of the Heisenberg model}
\label{sec:Heisenberg}

Next we examine the effect of the symmetries that are present in the spin-$1/2$ antiferromagnetic Heisenberg chain with $L$ sites. 
The system and the symmetry are discussed in many places, see {\it e.g.} \cite{furuya2017symmetry,Cheng:2015kce,Cho:2017fgz,Metlitski:2017fmd,Else:2019lft,Gioia:2021xtp,Cheng:2022sgb}.
There are two internal symmetries $G_x, G_y$, written in the computational basis of spins on each site: 
\be
G_x =X_1\otimes\dots\otimes X_L,\quad G_y=Y_1\otimes\dots\otimes Y_L.
\ee
We have further abused the notation to directly write the generator $G_i$ instead of its unitary representation $U(G_i)$. Choosing periodic boundary condition, we further impose the translation symmetry $T$, which maps site $i$ to site $(i+1)$ mod $L$. These symmetries satisfy the relations
\be 
G_x^2=G_y^2=1,\quad G_xG_y=(-1)^LG_yG_x,\quad T^L=1,\quad [G_x,T]=[G_y,T]=0.
\ee

To further see the mixed anomaly between translation and $G_i$, we introduce one defect $X_j$ at site $j$ acting on the spin chain as $\mathbbm{1}_1\otimes\dots\otimes \mathbbm{1}_{j-1} \otimes X_j\otimes \mathbbm{1}_{j+1} \otimes \dots\otimes \mathbbm{1}_L$. 
We define a new translation operator in the presence of the defect as a matrix product
\be
A:=X_jT. 
\ee
It satisfies the  property $A^2=X_jX_{j+1}T^2$, which we can use iteratively to derive $A^L=G_x$ and $A^{2L}=1$. Therefore, we can specify the symmetry group $\tilde{G}$ using only two generators $A$ and $G_y$:
\be
\tilde{G}=\langle A, G_y \mid  A^{2L}=1,\  G_y^2=1,\  AG_y=-G_y A \rangle.
\ee
Since the order-$2L$ element $A$ anticommutes with the $\mathbb{Z}_2$ generator $G_y$, and $-1$ is another order-$2$ element, the group is
$\tilde{G}=(\mathbb{Z}_{2L}\times \mathbb{Z}_2)\rtimes\mathbb{Z}_2$, where the generators for the three factors are $A,G_y,-1$, respectively. We will choose the density matrix $\rho$ to be of dimension $2^L$ and denote it in the presence of the defect $X_j$ as $\rho(X_j)$. Without loss of generality we choose to insert the defect at the first site $j=1$. 

To gain some intuition, we consider the simplest case of $L=2$. In the computational basis of local spins, $G_x=X_1\otimes X_2,\, G_y=Y_1\otimes Y_2,$ and
\be
T=\begin{pmatrix}
   1&0&0&0\\0&0&1&0\\0&1&0&0\\0&0&0&1
\end{pmatrix}\,\quad \Rightarrow  \quad A=\begin{pmatrix}
   0&1&0&0\\0&0&0&1\\1&0&0&0\\0&0&1&0
\end{pmatrix}.
\ee
The symmetry group $\left(\mathbb{Z}_4\times\mathbb{Z}_2\right)\rtimes\mathbb{Z}_2$ is SmallGroup$(16,3)$ in GAP. The only multi-dimensional representations are two-dimensional, so the whole spectrum must be twofold degenerate. More explicitly, we can find the similarity transformation $V$, such that
\be
A'=V^{-1}AV=\sigma_x\oplus i\sigma_y,\quad G_y'=V^{-1}G_yV=-\left(\sigma_z\oplus\sigma_z\right),\quad V=\begin{pmatrix}
    1&0&0&2\\0&1&-2&0\\0&1&2&0\\1&0&0&-2
\end{pmatrix}.
\ee 
$A'$ and $G_y'$ furnish a four-dimensional reducible representation of SmallGroup$(16,3)$, which decomposes into two different two-dimensional irreducible representations.\footnote{The polycyclic generating sequence (pcgs) \cite{hulpkedepartment} consists of f$1$, f$2$, f$3$, f$4$, with two two-dimensional irreducible representations in the GAP library:
\be
U^{(1)}:\quad\text{f}1=\sigma_x,\,\text{f}2=\sigma_z,\,\text{f}3=-\mathbbm{1}_2=-\text{f}4,
\ee
and 
\be
\label{eq:rep2}
U^{(2)}:\quad\text{f}1'=-i\sigma_y,\,\text{f}2'=\sigma_z,\,\text{f}3'=\text{f}4'=-\mathbbm{1}_2.
\ee
Here our first representation consists of $\tilde{U}^{(1)}(A)=\text{f}1\star\text{f}4,\,\tilde{U}^{(1)}(G_y)=\text{f}2\star\text{f}3$; the second consists of $\tilde{U}^{(2)}(A)=\text{f}1'\star\text{f}4',\,\tilde{U}^{(2)}(G_y)=\text{f}2'\star\text{f}3'$. Note they are not faithful.} In other words, $d_M=2$ for the average symmetry, and the symmetry is anomalous.

For general $L$, the order of $\tilde{G}$ is $8L$ with the group elements being\footnote{At $L=3$, $\tilde{G}$ is SmallGroup$(24,10)$, or $D_8\times \mathbb{Z}_3$. At $L=4$, $\tilde{G}$ is SmallGroup$(32,5)$, or $(\mathbb{Z}_8\times \mathbb{Z}_2) \rtimes \mathbb{Z}_2.$}
\be
\{1,A,A^2,\ \cdots,\ A^{2L-1},G_y, AG_y=-G_yA,\ A^2 G_y,\ A^3 G_y, \cdots A^{2L-1} G_y\}.
\ee
To obtain the minimal projective representation of $G$ that realizes the above extension $\tilde G$, we can construct the representation starting from an auxiliary vacuum $|0\rangle$ and acting on it the symmetry generators. If the resulting state can be distinguished from $|0\rangle$ by the eigenvalue of other symmetry generators due to the phases in the commutation relations of symmetry generators, the collection of such states including the vacuum forms a basis for the minimal projective representation.
In our case, the minimal projective representation can be described by
\begin{equation}
    \{|0\rangle,\ A|0\rangle\}~,
\end{equation}
where the two states are distinguished by the eigenvalue of $G_y$ due to $AG_y=-G_yA$.
We note that higher powers of $A$ acting on the above states do not give rise to new states distinguished by the eigenvalues of $G_y$.
Thus the dimension of the minimal projective representation is
\begin{equation}
    d_M=2~.
\end{equation}
The fact that the projective representation of $G$ has dimension $d_M$ greater than one again represents an anomaly of the average symmetry.

We would like to comment on the relationship of this example with the familiar disordered Heisenberg chain, where bond disorders drive a short-ranged antiferromagnetic Heisenberg chain into a random singlet phase \cite{PhysRevLett.48.344,PhysRevB.50.3799,IGLOI_2005}, whose entanglement entropy scales logarithmically \cite{PhysRevLett.93.260602,Refael_2009}. As will be discussed in section \ref{sec:ensemble}, in the cases of quenched disorders, the degeneracy in the enlarged density matrix arising from anomalous average symmetry is a degeneracy between the ground states of Hamiltonians with different disorder realizations. Therefore, our result does not contradict with the intuition of random singlet states. 
To constrain whether the system is gapped or gapless requires additional knowledge of the Hamiltonian. We will discuss more about this in
Section \ref{sec:discussion}.

\subsection{Example: toric code ground state}
\label{sec:toric}

In this section, we move on to the topologically  ordered systems and discuss the toric code model in (2+1)d \cite{Kitaev_2003} on space with torus topology. Putting qubits on the links of a square lattice on a torus, we are interested in the following microscopic symmetry generators, 
\be
U(g_1)= \prod_{l\in\gamma} \sigma^z_l,\quad U(g_2)=\prod_{l\in\tilde{\gamma}}  \sigma^x_l,
\label{eq:TC}
\ee
where $\gamma$ is a closed path on the lattice while $\tilde{\gamma}$ is a closed path on the dual lattice. These paths can wind around the two non-contractible cycles along the $x$- and $y$-directions on a 2d torus, giving $g_1^x$, $g_1^y,$ $g_2^x$, $g_2^y$.

We will now choose a basis and restrict to the 4-dimensional ground state subspace of the (2+1)d toric code model written in the computational basis of the logical qubits, 
$\{|00\rangle, |01\rangle, |10\rangle, |11\rangle\}$. The symmetries described in \eqref{eq:TC} act on the logical qubits as 
\be
U'(g_1^x)= Z \otimes \mathbbm{1},\quad 
U'(g_1^y)= \mathbbm{1}\otimes Z,\quad U'(g_2^x)= X \otimes \mathbbm{1},\quad 
U'(g_2^y)= \mathbbm{1}\otimes X.
\ee
These symmetries generate two decoupled copies of Pauli group. 
The dimension of the minimal representation is 
\begin{equation}
    d_M=2^2=4~.
\end{equation}
For a generic $4$-by-$4$ density matrix $\rho$, it is easy to prove following the logic in Appendix \ref{sec:Schur} that if $\rho$ satisfies 
\be
U'(g_a^i)\,\rho\, U'(g_a^i)^{-1} = \rho,\  \text{for}\ a=1,\ 2,\ \text{and}\ i=x,\ y,
\ee
then $\rho=\mathbbm{1}/4$. The take-home message is that a mixed state satisfying the mixed 1-form symmetries for toric code has to be an equal-weight maximally mixed state of the four ground states of toric code. This mixed state is separable, \textit{i.e.}, no additional entanglement is generated from such mixture. 

In the next section \ref{sec:entanglement}, we will move beyond the ground state subspace and also take into consideration the effect of defects.

\subsection{Anomalous average symmetry and topology of space
}

The projective action of the average symmetry generator on the Hilbert space can depend on the global topology of the space. As a consequence, the density matrix with anomalous average symmetry can subject to different constraints depending on the space topology.
 
 \paragraph{Example: toric code in 2+1d}

To illustrate this point, let us consider toric code model in (2+1)d \cite{Kitaev_2003} where each edge hosts a qubit, acted by Pauli matrices $X,Y,Z$. The model has $\mathbb{Z}_2$ symmetries generated by closed electric Wilson loops given product of Pauli $Z$ operators on the edges along the loop, and magnetic loops given by product of Pauli $X$ operators on the edges intersect the loop on the dual lattice \cite{Kitaev_2003}, see \eqref{eq:TC}.  The electric and magnetic loops have mutual $\pi$ statistics when they intersect an odd number of times.
We will consider the density matrix with these average symmetries.

Let us compare the symmetries when the space has different topologies.
\begin{itemize}
    \item 
When the space has topology of a sphere of trivial intersection form, there is no non-contractible cycles. Thus the average symmetry generators always commute, and the average symmetries are not anomalous. The dimension of the minimal projective representation is $d_M=1$.

\item
On the other hand, when the space has topology of a torus, which has first homology $H_1(T^2)=\mathbb{Z}_2\times\mathbb{Z}$ with nontrivial intersection form,
the average symmetry generators can obey anomalous commutation relations due to the $\pi$ statistics.
The dimension of the minimal projective representation is $d_M=2^2=4$. Thus the eigenvalues of the density matrix with average symmetry has degeneracy of at least $4$.

\item When the space has topology of a closed orientable Riemann surface with genus $n$, the first homology is generated by the one-cycles $\{\alpha_i,\beta_i\}$ for $i=1,\cdots,n$, and $\#(\alpha_i,\beta_j)=\delta_{ij}$, while $\{\alpha_i\},\{\beta_i\}$ separately do not intersect among themselves. The minimal representation can be constructed as
\begin{equation}
\left\{  
 Z\left(
 \prod_i
 \alpha_i^{k_i}
 \beta_i^{m_i}\right)
 |0\rangle
 \right\}~,
\end{equation}
for some vacuum state $|0\rangle$, and $k_i,m_i=0,1$. These states can be distinguished by the eigenvalues of the magnetic loops that intersect the above electric loop.
The dimension of the minimal representation is
\begin{equation}
    d_M=4^n~.
\end{equation}
Thus the density matrix with the average symmetry given by the loop operators have degenerate spectrum with degeneracy of at least $4^n$.

\end{itemize}

\subsection{Symmetry in ensembles with random couplings}
\label{sec:ensemble}

In this section, we will study ensembles labelled by a random coupling $\lambda$ with  probability $\mu(\lambda)\geq 0$.
Typical example include 
classical and quantum spin glasses \cite{RevModPhys.58.801,Castellani_2005,cugliandolo2022quantum}, 
Sachdev--Ye--Kitaev (SYK) models \cite{Sachdev:2015efa,kitaevTalks,kitaevkitp1,kitaevkitp,Maldacena:2016hyu,Kitaev:2017awl,Qi:2021oni}, and they play an important role in understanding the holographic correspondence for replica wormholes in quantum gravity.
For recent discussions, see {\it e.g.}  \cite{Hsin:2020mfa,Chen:2020ojn,Belin:2020hea,Saad:2021rcu,Antinucci:2023uzq}.
The random coupling $\lambda$ can also label different configurations of quenched disorder.

Let us consider average symmetry that is violated in each member of the ensemble with fixed a random coupling $\lambda$.
The density matrix $\rho_\lambda$ of a given coupling $\lambda$ satisfies
\begin{equation}\label{eqn:avgsymensemble}
    U_\lambda \rho_\lambda U_\lambda^{-1}=\rho_{f(\lambda)}~,
\end{equation}
where $f$ is an outer automorphism on the random couplings: $f$ maps a coupling uniquely to another coupling.
The operator $U$ is not an average symmetry of $\rho_\lambda$, since it maps the density matrix to a different density matrix for $f(\lambda)\neq \lambda$.

We can make $U_\lambda$ an average symmetry of an enlarged density matrix. Let us enlarge the Hilbert space ${\cal H}=\oplus_\lambda {\cal H}_\lambda$ where ${\cal H}_\lambda$ is the Hilbert space for a given random coupling $\lambda$.
We introduce ``ancilla" $|\lambda\rangle$ that labels the random coupling $\lambda$, with $\langle \lambda |\lambda'\rangle=\delta_{\lambda,\lambda'}$.
The enlarged density matrix is
\begin{equation}\label{eqn:ensembledensitymatrix}
    \rho=\sum_\lambda \mu(\lambda) \rho_\lambda\otimes |\lambda\rangle \langle \lambda|~.
\end{equation}
The average expectation value of an operator ${\cal O}$ in the ensemble is
\begin{equation}
    \overline{\langle {\cal O} \rangle}=\text{Tr}(\rho {\cal O})=\sum_\lambda \mu(\lambda) \text{Tr}_{{\cal H}_\lambda}\left(\rho_\lambda {\cal O}\right)~.
\end{equation}
When the random couplings are continuous real numbers, the sum over couplings is replaced by an integral, and $\langle \lambda|\lambda'\rangle=\delta(\lambda-\lambda')$.

When the distribution of the random coupling satisfies $\mu(f(\lambda))=\mu(\lambda)$, the average symmetry of the ensemble corresponds to $U_\lambda$ in (\ref{eqn:avgsymensemble}) is
\begin{equation}\label{eqn:enlargedsymmetry}
    U=\sum_\lambda U_\lambda\otimes |f(\lambda)\rangle \langle \lambda|~.
\end{equation}
To see this, let us verify $U\rho=\rho U$:
\begin{equation}
    U\rho=\sum_{\lambda',\lambda} \mu(\lambda) U_{\lambda'} \rho_{\lambda}|f(\lambda')\rangle\langle \lambda' |\lambda\rangle \langle \lambda| 
    =\sum_{\lambda} \mu(\lambda) U_{\lambda}\rho_{\lambda}|f(\lambda)\rangle\langle \lambda|~.
\end{equation}
On the other hand,
\begin{equation}
    \rho U=\sum_{\lambda',\lambda}\mu(\lambda')
    \rho_{\lambda'} U_{\lambda} |\lambda'\rangle \langle \lambda'| f(\lambda)\rangle \langle \lambda|
    =\sum_{\lambda}\mu(f(\lambda)) \rho_{f(\lambda)} U_{\lambda}|f(\lambda)\rangle \langle \lambda|~.
\end{equation}
Using $\mu(f(\lambda))=\mu(\lambda)$ and $U_\lambda \rho_\lambda=\rho_{f(\lambda)}U_\lambda$, we find $U\rho=\rho U$, {\it i.e.} $U$ is an average symmetry of $\rho$.

Therefore we can apply the constrain from anomalous average symmetry on the enlarged density matrix (\ref{eqn:ensembledensitymatrix}). 
There are several important differences from the previous non-ensemble cases:
\begin{itemize}
    \item The symmetry acting on the enlarged density matrix in general does not factorize to the operator acting on a region and its complement $U\neq U_AU_B$ due to the sum over $\lambda$, the reduced density matrix in general does not have average symmetry. As a consequence, the entanglement entropy of the enlarged density matrix is not directly constrained as we will discuss in Section \ref{sec:entanglement}. 

    \item 
Here we require the symmetry generators to act on the ancilla as permutations corresponding to the maps $f$ for the symmetries.
In particular, since $\mu(f(\lambda))=\mu(\lambda)$, the part of density matrix acting on the ancilla $\mu(\lambda)|\lambda\rangle\langle \lambda|$ automatically has degenerate eigenvalues due to the symmetry, regardless of whether the symmetry is anomalous. 

\end{itemize}

Denote the eigenvalues of $\rho_\lambda$ by $r^i_\lambda$, where $i=1,2,\cdots, N_\lambda$ with $N_\lambda$ being the dimension of the Hilbert space ${\cal H}_\lambda$.
Then the eigenvalues of $\rho$ are
$\{r^i_\lambda \mu(\lambda)\}$ for different $\lambda$.
Suppose the anomalous average symmetry of $\rho$ has dimension $d_M$ for the minimal projective representation with symmetry generators acting on the ancilla by the given permutations $f$. The set of symmetries generators permute the couplings $\lambda$ with orbit of length $\ell$.
Then the eigenvalues of $\rho_\lambda$ have degeneracy
\begin{equation}
    d_M'=d_M/\ell~.
\end{equation}

For instance, if the average symmetry does not permute the couplings, $\ell=1$, and this means $\rho_\lambda$ itself has anomalous average symmetry that gives eigenvalue degeneracy $d_M$. 

As another simple example, suppose there are two couplings with equal probability, and the enlarged density matrix is $(1/6)\text{diag}(1,2,2,1)$, where $\rho_\lambda=(1/3)\text{diag}(1,2)$ and $(1/3)\text{diag}(2,1)$ for the two couplings, respectively.
The average symmetry generators are $U_1=X\otimes \text{SWAP},\ U_2=Z\otimes \mathbbm{1}$ where SWAP exchanges the two couplings while $\mathbbm{1}$ does not. The average symmetry is anomalous since $XZ=-XZ$, which gives the minimal dimension
\begin{equation}
    d_M=2~.
\end{equation}
This is consistent with the double degeneracy of eigenvalues for the enlarged density matrix $\rho$.
On the other hand, the average symmetry permutes the two couplings with orbit length $\ell=2$. Thus each $\rho_\lambda$ has eigenvalue degeneracy 
\begin{equation}
 d_M'=2/2=1~.   
\end{equation}

 We remark that the role of ancilla is similar to spurion that restores an explicitly broken global symmetry (see {\it e.g.} \cite{Komargodski:2011xv} for dilaton as the spurion for conformal symmetry).

\section{Application: Bounds on Entanglement Entropy}
\label{sec:entanglement}

Entanglement entropy plays an important role in distinguishing phases of matter and phase transitions, with the capability of being experimentally measured  \cite{Islam:2015mom}. For instance, the entanglement entropy of a contractible subregion in the ground states of gapped phases is believed to scale with the area of the subregion instead of the volume (see \cite{Hastings:2007iok,Eisert:2008ur} and the references therein). 
It is also used as the most important characterization of entanglement phase transitions in monitored quantum circuits, see for example \cite{circuit_review}. We will explore the above two branches and their interplay with anomalous average symmetries later in this section. Other significance of entanglement also include, but not limited to,
the extraction of properties of criticality such as the conformal central charge in (1+1)d (see {\it e.g.} \cite{Calabrese:2009qy,Vanhove:2021zop,Huang:2021nvb, bengtsson2017geometry,rangamani2017holographic}. 
Additionally, entanglement also manifests its importance in the context of holographic duality \cite{Maldacena_1999,Witten:1998qj,Aharony_2000}, especially in the Ryu-Takayanagi formula \cite{RT}.

In this section, we will apply the constraints on the density operator discussed in Section \ref{sec:densitymatrix} to the scaling of entanglement entropy. We divide the space into two regions $A$ and its complement $B=A^c$. Starting from a density matrix $\rho$ (which can be a pure state), we define the reduced density matrix $\rho_A$ on region $A$ by taking trace with respect to states supported on region $B$ (recall we assume the Hilbert space is a tensor product of local Hilbert space of finite dimension), then the entanglement entropy is
\begin{equation}
    S(\rho_A)=-\text{Tr}\left(\rho_A\log\rho_A\right)
    =-\sum_{\lambda_i\neq 0} \lambda_{i}\log\lambda_i=-\sum_{\text{distinct }\lambda\neq 0}d(\lambda)\lambda\log\lambda~,
\end{equation}
where $\lambda_i> 0$ are the nonzero eigenvalues of the reduced density operator $\rho_A$ (the zero eigenvalues do not contribute to $S(\rho_A)$), and $\lambda$ are the distinct nonzero eigenvalues with degeneracy $d(\lambda)$.

Similarly, we can derive a lower bound on the $\alpha$-R{\'e}nyi entropy $S_\alpha(\rho_A)$ for all real $\alpha>1$, where the $\alpha$-R{\'e}nyi entropy is
\begin{equation}
    S_\alpha(\rho_A)=-\frac{1}{\alpha-1}\log\text{Tr}\left(\rho_A^\alpha\right)
    =-\frac{1}{\alpha-1}\log\sum_{\lambda_i\neq 0} \lambda_i^\alpha=-\frac{1}{\alpha-1}\log\sum_{\text{distinct }\lambda\neq 0}d(\lambda)\lambda^\alpha
    ~.
\end{equation}

\subsection{Average symmetry of reduced density matrix}

Let us constrain the reduced density matrix $\rho_A$ using the symmetry of the original density matrix $\rho$. Suppose $\rho$ has average symmetry
\begin{equation}\label{eqn:rhosymm}
    U\rho U^{-1}=\rho~.
\end{equation}
For instance, if the Hamiltonian has exact symmetry $G$, an example of density matrix with average symmetry $G$ is 
\begin{equation}
    \alpha \sum_{g\in G} |g\Psi\rangle\ \langle g\Psi|~,
\end{equation}
where $\Psi$ is a state with given energy, and $g\Psi$ is the state transformed by the symmetry generator for $g\in G$, and it has the same energy as $\Psi$. The overall normalization $\alpha$ is fixed by the unit trace condition.

The Hilbert space is the tensor product of local Hilbert spaces, and we consider symmetry that factorizes as
\begin{equation}
    U=U_AU_B~,
\end{equation}
where $U_A$ acts only on region $A$, and $U_B$ acts only on the complement region $B$. 
Such property is called splittability (see \cite{Harlow:2018tng} and the references therein).
Both $U_A,U_B$ are invertible, since $U$ is invertible.
By suitable scaling with an overall factor, We can take $U_A,\ U_B$ to be both unitary operators for unitary $U$.
For instance, if the symmetry is on-site, then the above factorization property holds for any region $A$. This is not always the case, such as the situation discussed in Section \ref{sec:ensemble}. The condition represents a strict locality of the symmetry action, and it would be interesting to generalize the argument to other cases. For now, we will consider the region $A$ where the condition holds.

Denote $\{|b\rangle\}$ to be a basis for the Hilbert space on region $B$. Taking trace over such states on both sides of the relation (\ref{eqn:rhosymm}) and using $U=U_AU_B$ gives
\begin{equation}
    U_A\left(\sum_b \langle b| U_B \rho U_B^{-1}|b\rangle\right)U_A^{-1}
    =\sum_b
    \langle b| \rho|b\rangle~.
\end{equation}
The left hand side can be simplified as follows: we note that since $U_B$ is unitary, $\{U_B^{-1}|b\rangle\}=\{|b'\rangle\}$ forms another basis of the Hilbert space supported on region $B$. In particular, if the states $\{|b\rangle\}$ are orthonormal, the states $\{|b'\rangle\}$ are also orthonormal. Thus the reduced density matrix also has average symmetry
\begin{equation}
    U_A \rho_A U_A^{-1}=\rho_A~,
\end{equation}
where 
\begin{equation}
    \rho_A=\sum_b \langle b|\rho|b\rangle=\sum_{b'} \langle b'|\rho|b'\rangle~.
\end{equation}
Therefore we can use the symmetry action on a subregion to constrain the reduced density matrix in that region. Several remarks are in place: 
\begin{itemize}
    \item 
The symmetry acting on a subregion $A$ can depend on the region geometry.
For instance, if there is average symmetry localized on each site, then the symmetry acting on the region has order that grows with the system size, and the constraint on the reduced density matrix depends on the volume of the region.

\item
If we replace average symmetry by exact symmetry, the argument would not go through. The density matrix $\rho$ we start with can have a strong symmetry, but the reduced density matrix in general does not have corresponding strong symmetry.

\item 
For the cases discussed in subsection \ref{sec:ensemble}, where the average symmetry is realized in an ensemble due to some random terms in the Hamiltonian/Lagrangian, the correct calculation of entanglement entropy should be to first calculate the entanglement entropy for each pure state which arises from one realization of the random term, and then average over different realizations. This is different from computing the entanglement entropy directly from the density matrix, {\it i.e.} first averaging over realizations and then calculate the entanglement entropy. In other words, the average symmetry that acts on the density matrix generically will not factorize and the reduced density matrix has no average symmetry.
\end{itemize}

\subsection{Anomalous symmetry gives lower bound on entanglement entropy}

As discussed in Section \ref{sec:densitymatrix}, anomalous average symmetries of the system constrains the density operator to be ``close to the identity matrix'': the spectrum of the density operator has degeneracy bounded from below by the dimension $d_M>1$ of the minimal representation that realizes the nontrivial phases in the anomalous commutation relations. In the following we will derive a bound on the entanglement entropy for systems with anomalous average symmetry.

Let us first show that the eigenvalues of the density operator are sufficiently small, such that the contribution of each nonzero eigenvalue to the density operator is positive $-\lambda \log\lambda>0$.
To see this, we can use the unit trace condition
\begin{equation}
    1=\text{Tr}(\rho_A)=\sum_{\lambda} d(\lambda)\lambda\geq \sum_{\lambda} d_M \lambda\quad \Rightarrow\quad  \sum_{\lambda} \lambda \leq \frac{1}{d_M}<1~,
\end{equation}
where we used $d(\lambda)\geq d_M$ and $d_M>1$.
Since the eigenvalues $\lambda$ are non-negative, the last equation implies that every nonzero eigenvalue $\lambda$ satisfies
\begin{equation}
    \lambda \leq \frac{1}{d_M}~\Rightarrow -\log\lambda \geq \log d_M~.
\end{equation} 
The entanglement entropy is therefore bounded from below as
\begin{equation}
    S(\rho_A)=-\sum_{\lambda} d(\lambda)\lambda\log\lambda \geq  \sum_{\lambda} d(\lambda)\lambda \cdot \log d_M=\log d_M~,
\end{equation}
where we have again used the unit trace condition. When the average symmetry enlarges with the system size, such that $d_M$ also grow with the size of the region, the growth of the entanglement entropy is bounded from below by the growth of $\log d_M$. As an extreme example, if the reduced density operator is proportional to the identity matrix, 
\be
\rho_A=\frac{1}{\text{dim }{\cal H}}\mathbbm{1}= \mathbbm{1}/{n^V}
\ee
where $V$ is the volume of the region $A$ and $n$ is the dimension of the local Hilbert space. Then the entanglement grows with the volume of the region, $S(\rho_A)=V\log n$.

Similarly, the $\alpha$-R{\'e}nyi entropy of the reduced density matrix $\rho_A$ for real $\alpha>1$ is bounded from below by
\begin{equation}
    S_\alpha(\rho_A)=-\frac{1}{\alpha-1}\log \sum_i\left( \lambda_i\cdot \lambda_i^{\alpha-1}\right)
    \geq 
    -\frac{1}{\alpha-1}\log \sum_i  \left(\lambda_i/d_M^{\alpha-1}\right)
    =\log d_M~.
\end{equation}
This is a stronger bound than the entanglement entropy since $S_\alpha>S_{\alpha'}$ for $\alpha<\alpha'$ (see {\it e.g.} \cite{Muller-Lennert:2013liu}). The R{\'e}nyi entropy for the reduced density matrix in the limit $\alpha\rightarrow 1^+$ gives the entanglement entropy.

\subsection{Example: obstruction to many-body localization (MBL)}

Many-body localization (MBL) describes a class of systems that do not approach thermal equilibrium \cite{Gornyi_2005,Basko_2006}. For a review, see {\it e.g.} \cite{Nandkishore_2015,RevModPhys.91.021001}.
Such systems have local symmetries called l-bits (where l stands for localized).
For instance, if the l-bits are Pauli operators $\{Z_i\}$ at each site $i$, 
the Hamiltonian can be written as
\begin{equation}
    H_\text{MBL}=E_0+\sum Z_i+\sum J_{ij}Z_iZ_j+\cdots~.
\end{equation}
The Hamiltonian commutes with the local symmetry generators $\{Z_i\}$. These symmetries are supported at points, and they are are $(D-1)$-form symmetry in $D$ spacetime dimension \cite{Gaiotto:2014kfa}. 
In the following, we will show how anomalous symmetry can constrain the existence of many body localization. 

While many-body localized states are typically pure eigenstates of an MBL Hamiltonian, here we will be interested in mixed localized states, which can be obtained from the pure states by coupling to the environment or raising the temperature. 
We will consider the states with localized average symmetries, and investigate the entanglement entropy in a contractible region. We will consider the situation that the symmetry generators acting on the contractible region do not commute, thus the reduced density matrix has anomalous average symmetry. For simplicity, we consider the example where each site $i$ can host both the $Z_i$ and $X_i$ symmetries which do not commute.
Then the minimal dimension is at least 
\begin{equation}
 d_M\geq 2^{V}~,   
\end{equation}
where $V$ is the ($(D-1)$-dimensional) volume of the region.

From the previous discussion, the entanglement entropy for the reduced density matrix thus satisfies
\begin{equation}
    S(\rho_A)\geq \log d_M\geq V\log 2~.
\end{equation}
In other words, the entanglement entropy grows with the volume of the system, 
which contradicts with the expectation that a many-body localized state should be short-range entangled.
Therefore, the system cannot be in the many-body localized phase. This is consistent with the obstruction to many body localization  discussed in \cite{PhysRevB.94.224206,Parameswaran_2018,Gopalakrishnan_2020}, when non-abelian (or in our language anomalous) symmetries are present.

\paragraph{Example: disordered Heisenberg chain} 

 In particular for the disordered Heisenberg chain with $SO(3)$ symmetry that acts projectively as $SU(2)$ at each site, ref. \cite{PhysRevB.96.041122} presented an argument from the entanglement point of view, while ref. \cite{PhysRevB.94.224206} pointed out a general paradox without referring to entanglement. Here, we give an alternative argument that shows the symmetry prevents the many body localization phase.

\subsection{Example: area law entanglement of topological order}

Let us consider toric code lattice model in (2+1)d.
Each edge has a local qubit acted by Pauli $X,Y,Z$ operators. The Hamiltonian is
\begin{equation}
    H_0=-\sum_v \prod_{e:v\in \partial e} X_e-\sum_f \prod_{e\in\partial f} Z_e~.
\end{equation}
Excited states that are penalized by the first term carry electric charge. As reviewed also in Section \ref{sec:densitymatrix}, the theory has 1-form symmetries given by closed loops of the Pauli $X$ and the closed loops of the Pauli $Z$ operators
\begin{equation}
    Z(\gamma)=\prod_{e\in \gamma}Z_e,\quad X(\tilde \gamma)=\prod_{e\in \tilde\gamma}X_e~,
\label{eq:symmetry}
\end{equation}
where $\gamma$ is a loop on the lattice, and $\tilde \gamma$ is a loop on the dual lattice, and $e\in\tilde \gamma$ means the edge $e$ belongs to the edges corresponding to $\tilde \gamma$. 
Ground states of the Hamiltonian $H_0$ exhibit the symmetries \eqref{eq:symmetry} for contractible loops $\gamma$ and $\tilde{\gamma}$, and non-contractible symmetry operators can exchange the degenerate ground states. 
We will focus on the symmetries on contractible loops in the following discussion, and the discussion is not sensitive to global topology of the space.
Generically, the toric code phase may not arise from an exactly solvable model $H_0$, but we still expect these 1-form symmetries to emerge at low energies.

We will examine the entanglement entropy for a contractible disk region $A$ for a ground state. There can be multiple ground states depending on the topology of space, and our argument works for any ground state. Consider one-form symmetries supported on contractible loops across both the region $A$ and its complement region $B$. The symmetry satisfies the factorization condition $U=U_AU_B$, where $U_A,\ U_B$ can be open strings that concatenate into a closed loop. The symmetry $U$ is an exact symmetry of the ground state density matrix, and also an averaged symmetry.

\begin{figure}[t]
    \centering
   \includegraphics[width=0.5\textwidth]{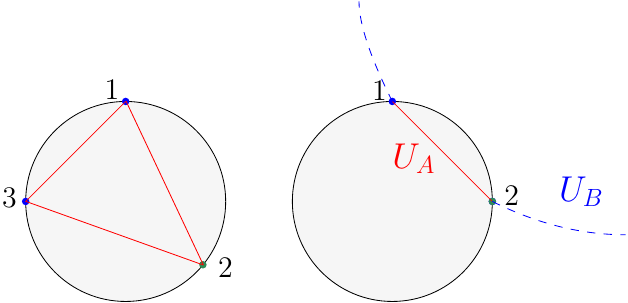}
    \caption{Symmetry generators acting on the grey shaded region $A$. In the left figure, the generators that are contractible within the region are trivial. In the right figure, the generators can be labelled by a point 2 on the circumference for a fixed reference point 1.}
    \label{fig:2dsymmetrytoriccode}
\end{figure}

Let us study the minimal representations of the symmetry $U_A$ in region $A$.
For each two points $1,2$ on the boundary of region $A$, there is a symmetry generator $U_A=Z_{12}$ on the string connecting the two points. There can be multiple open-string operators connecting the designated two endpoints, but they are all equivalent to each other, differ by local Hamiltonian terms supported in region $A$. On the other hand, the symmetry generators on a closed loop connecting three points on the boundary of $A$ can be contracted in region $A$ and represents trivial symmetry (see the left of Figure \ref{fig:2dsymmetrytoriccode}). Therefore it is sufficient to fix a reference point on the boundary $\partial A$, and the symmetry generators are on the strings connecting the reference point to another point $p$ on the boundary $\partial A$ (see the right of Figure \ref{fig:2dsymmetrytoriccode}). All of the symmetry generators given by the Pauli $Z$ strings can be distinguished by their commutation relations with different Pauli $X$ strings. If the length of the boundary $\partial A$ is $L$ in the lattice size unit, 
there are $(L-1)$ distinct strings for different points $p$ that do not coincide with the reference point.
Thus the minimal representation of the symmetry is generated by
\begin{equation}
    \left\{\;\prod_{p=1}^{L-1} Z_{p}^{n_p}|0\rangle,\; n_p=0,1\right\}~,
\end{equation}
where $|0\rangle$ is an auxiliary vacuum state, and $Z_p$ is the Pauli $Z$ string on the string connecting the reference point and $p$ on the boundary $\partial A$.
The dimension of the minimal representation is 
\begin{equation}
    d_M=2^{L-1}~.
\end{equation}
Thus the entanglement entropy for ground state on the region $A$ of boundary size $L$ satisfies
\begin{equation}\label{eqn:arealawtoriccode}
    S(\rho_A)\geq (L-1)\log 2=L\log 2 -\log 2~.
\end{equation}
The bound is saturated by topological order 
where $L$ is in the microscopic lattice constant units,
but it still holds if we deform the topological order to be a non-topological system, as long as the density matrix has
emergent or exact
one-form symmetries given by Pauli $X,Z$ strings, 
and then $L$ is now in the unit of length scale where the symmetries emerge.

We remark that the bound (\ref{eqn:arealawtoriccode}) not only constrains the universal constant part in the entanglement entropy $(-\log 2)$, but also constrains the area law entanglement coefficient for such lattice regularization. The constant part $(-\log 2)$ coincides with the topological entanglement entropy of the toric code ground state \cite{Levin_2006,Kitaev_2006}.\footnote{
We remark that the counting of entanglement entropy is reminiscent to the ``bit thread'' proposal for area-law holography entanglement entropy \cite{Freedman:2016zud}, but the discussion here is not related to the proposal there. In particular, the ``bit thread'' is not related to global symmetries.
}

\subsection{Example: toric code with charge condensation on sublattice and entanglement phase transition}

If we deform the system such that the symmetries become different, the bound on the entanglement can also change, and the system can undergoes a phase transition characterized by different entanglement properties as we increase the deformation.
We will consider the entanglement entropy for the ground state, and symmetries for both the Hamiltonian as well as density matrices.

Let us deform the (2+1)d toric code model with charge condensation defects on a sublattice $\Lambda$, where we remove the vertex terms for the vertices on $\Lambda$:
\begin{equation}
    H_\Lambda=-\sum_{v\in \Lambda^c} \prod_{e:v\in \partial e} X_e-\sum_f \prod_{e\in\partial f} Z_e~,
\end{equation}
where $\Lambda^c$ is the complement of $\Lambda$. As we will discuss below, the anomalous symmetries of the system depends crucially on the choice of $\Lambda$.
For instance, if $\Lambda$ is the entire lattice, $Z$ on each edge will commute with the Hamiltonian and thus the number of symmetry generator grows with the volume (in two space dimensions it is the area) of the system.

\paragraph{Alternating column condensation}

Let us condense the electric charge on alternating columns on the square lattice.
Consider the region $A$ to be a square of size $L_x\times L_y$. Without loss of generality we can take $L_x$ to be even.
There are new symmetries given by the strings with endpoint $p$ not on the boundary $\partial A$, but on one of the $L_x/2$ columns of length $L_y$.
Thus the dimension of the minimal representation is
\begin{equation}
    d_M=2^{L-1+L_y\frac{L_x}{2}}~,
\end{equation}
where $L=2(L_x+L_y)$ is again the size of the boundary $\partial A$.
The entanglement entropy on region $A$ is thus bounded by
\begin{equation}\label{eqn:chargecondensetoricocde}
    S(\rho_A)\geq \frac{L_xL_y}{2}\log 2+L\log 2-\log 2~.
\end{equation}
In particular, the entanglement entropy obeys volume law for such deformation.

\paragraph{Alternating site condensation}
Similarly, since the square lattice is bipartite, one can condense the electric charges on one of the two sublattices.
There are new symmetries by choosing the endpoint $p$ of the string to be on the condensed site. Since there are $L_xL_y/2$ such sites, the bound on the entanglement entropy is the same as (\ref{eqn:chargecondensetoricocde}).

\paragraph{General charge condensed sublattice}

Suppose within the region $A$, the ration of the condensed sites is $0<r<1$. Then there are new symmetries from strings whose endpoint belong to one of the condensed site.
Thus the dimension of the minimal representation is
\begin{equation}
    d_M=2^{L-1+r a}~,
\end{equation}
where $a$ is the size of the region $A$.
Thus the entanglement entropy is bounded by
\begin{equation}
    S(\rho_A)\geq  r a\log 2 +L\log 2-\log 2~.
\end{equation}

\paragraph{Entanglement transition}

As we deform the toric code model by increasing $r$ from 0 to 1, {\it i.e.} by changing the ratio of charge condensed sites,
the entanglement entropy changes from the area law (where the bound is saturated) to (at least) the volume law. The system undergoes a transition characterized by the change of the entanglement entropy scaling law.

\section{Application: Quantum Channels}
\label{sec:quantumchannel}

Consider the quantum channel in the Kraus representation \cite{kraus1983states,nielsen2010quantum}
\begin{equation}
 {\cal E}[\rho]=\sum_i K_i\rho K_i^\dag~,   
\end{equation}
where $\{K_i\}$ are the Kraus operators that satisfy $\sum_i K_i K_i^\dag=1$ to preserve the trace of the density matrix.
We note that the Kraus operators related by the following transformation represent the same quantum channel: (see {\it e.g.} \cite{nielsen2010quantum})
\begin{equation}
    K_j\rightarrow \sum_l u_{jl} K_{l}~,
\end{equation}
where $\mathbf{u}=(u_{ij})$ is a constant unitary matrix multiplied by the identity operator. In a basis of Kraus operator $\{K_i\}$ that diagonalize the unitary matrix constant, the condition can be written as
\begin{equation}
    K_j\rightarrow e^{i\phi_j}K_{j}~,
\end{equation}
where $e^{i\phi_j}$ is a phase for each Kraus operator $K_j$.

We will discuss two kinds of symmetry constraints on the quantum channels. In the first kind, the transformed density matrix ${\cal E}[\rho]$ has the average symmetry for any initial density matrix $\rho$, and thus the symmetry only depends on the quantum channel. In the second kind, the initial density matrix $\rho$ has an average symmetry preserved by the quantum channel, {\it i.e.} the transformed density matrix ${\cal E}[\rho]$ also has the same average symmetry of $\rho$.

\subsection{Average symmetry from quantum channel}

We consider the quantum channel with symmetry specified by generators $U$ that act on the Kraus operators as a unitary representation
\begin{equation}\label{eqn:symmetryKraus}
    UK_j= \sum_{l} u_{jl}K_{l}~,
\end{equation}
where the matrix $(u_{jl})$ is a constant unitary matrix multiplied the identity operator.
In a basis of Kraus operators where the unitary matrix $(u_{ij})$ is diagonalized, it can be written as
\begin{equation}\label{eqn:symmetryKrausp}
    UK_j= e^{i\phi_j }K_{j}~,
\end{equation}
where $e^{i\phi_j }$ is a phase. 
Then the resulting density matrix has average symmetry $U$ for any initial density matrix:
\begin{equation}
    U{\cal E}[\rho]U^{-1}=
    \sum_i UK_i\rho (UK_i)^\dag=\sum_{i,j,l} u_{ij}K_{j}\rho K_l^\dag u_{il}^*={\cal E}[\rho]~,
\end{equation}
where the last equality used $\sum_i u_{il}^*u_{ij}=\delta_{jl}$ and $(u_{ij})$ being a constant unitary matrix multiplied the identity operator.
Thus the symmetry $U$ is an intrinsic symmetry property of the quantum channel itself, independent of the initial density matrix $\rho$.

When the symmetry is anomalous with dimension $d_M$ for the minimal projective representation, the quantum channels with symmetry in this case have at least $d_M$ Kraus operators. Furthermore, the discussion in Section \ref{sec:entanglement} implies that such quantum channel with anomalous symmetry must produce states with nontrivial von Neumann entropy (and entanglement entropy, depending on the subregion).

We note that the situation discussed here is also discussed in \cite{de_Groot_2022}, where each Kraus operator is invariant under conjugating by the symmetry generator up to a phase.

\paragraph{Example: the depolarization channel}

Consider the example of decoherence given by the following depolarization channel. Each site has a qubit, and the channel at site $i$ transforms the density matrix as
\begin{equation}
    {\cal E}_i[\rho]=(1-p)\rho+\frac{p}{3}\left(X_i \rho X_i+Y_i\rho Y_i+Z_i\rho Z_i\right)~,
    \label{eq:decoherence}
\end{equation}
where $p$ is a real number between 0 and 1.
The total quantum channel is given by composing the channels on all the sites. At $p=0.75$, the Kraus operators are the Pauli matrices with the same coefficients, and the quantum channel has Pauli group symmetry.
The Pauli group average symmetry at each site is anomalous, with
\begin{equation}
    \text{Each site}:\quad d_M=2~.
\end{equation}
In fact, the channel in this case produces the maximally mixed state for any initial density matrix (see {\it e.g.} \cite{nielsen2010quantum}). 
Thus the dimension of the minimal projective representation is $d_M=2^V$ where $V$ is the volume of the system.
The entanglement entropy for any region $A$ satisfies volume law
\begin{equation}
    S(\rho_A)\geq V(A)\log 2~,
\end{equation}
where $V(A)$ is the volume of the region $A$. In other words, the quantum channel with anomalous symmetry creates volume law entanglement regardless of the initial density matrix.

On the other hand, the channel with $p=0$ is the identity channel that does not transform the density matrix. Thus if we start with a density matrix with area law entanglement entropy and apply the quantum channel, as we dial $p$ from 0 to 0.75, the system undergoes a transition in the entanglement entropy from the area law to volume law.

We would like to point out that \eqref{eq:decoherence} can also be obtained from a monitored circuits with weak measurements, with the following setup: Starting from the initial state $\rho_0$, at each time step, randomly pick a site and conduct weak measurements on it.  On the chosen site, there is probability $2p/3$ to measure $X_i$, $2p/3$ to measure $Y_i$, probability $2p/3$ to measure $Z_i$, and probability $(1-2p)$ to do nothing. The dynamics of the system is then described by
\be
\rho(t+1)=
\frac{1}{N}\sum_i \left(\frac{2p}{3} \sum_{a=x,y,z} \sum_{s_i^a=\pm 1} \frac{1+s_i^a \sigma_i^a}{2}\rho \frac{1+s_i^a \sigma_i^a}{2}+(1-2p)\rho\right)~.
\ee
The parenthesis on the right hand side coincides with the quantum channel (\ref{eq:decoherence}):
\be
(1-2p)\rho+\frac{2p}{3} \sum_a \frac{\rho + \sigma_i^a \rho \sigma_i^a}{2}=(1-p)\rho+\frac{p}{3}\sum_a \sigma_i^a\rho\sigma_i^a={\cal E}_i[\rho(t)]~.
\ee

\subsubsection{``Emergent'' average symmetry from iterative quantum channel} 

We note that applying quantum channel $N$ times gives the effective channel:
\begin{equation}
    {\cal E}^N[\rho]=\sum_{i_1,i_2,\cdots, i_N} K_{i_N}\cdots K_{i_2} K_{i_1}\rho \left(K_{i_N}\cdots K_{i_2} K_{i_1}\right)^\dag:=\sum_I K_I \rho K_I^\dag~, 
\end{equation}
where the effective Kraus operators are $K_I:=K_{i_N}\cdots K_{i_1}$, where $I$ is the collective indices $I=(i_1,\cdots ,i_N)$.
There can be symmetry from the iterating quantum channel ${\cal E}^N$ for some sufficiently large $N$ but not for single channel $N=1$.

Consider the large iteration channel ${\cal E}_q^N$ of the following channel with general probability parameters $p_1,p_2,p_3$:
\begin{equation}
    \mathcal{E}_g[\rho]=q\rho+p_1X\rho X+p_2Y\rho Y+p_3Z\rho Z~,
\end{equation}
which satisfy $q=1-p_1-p_2-p_3$. The real numbers $p_1,p_2,p_3$ are between 0 and 1. 
For convenience, denote by $x,y,z$ the conjugations by the Pauli $X,Y,Z$ operators, respectively. Keep in mind that $x,y,z$ are nontrivial elements of the Klein four-group. Then each iteration of applying $\mathcal{E}_g$ is equivalent to multiplying by $(q+p_1\,x+p_2\,y+p_3\,z)$. Suppose at the $N^{\text{th}}$ iteration, the resulting channel is 
\begin{equation}
    \mathcal{E}_g^N[\rho]=f\rho+gX\rho X+hY\rho Y+kZ\rho Z~,
\end{equation}
where $f,g,h,k$ are degree-$N$ polynomials in $q,p_1,p_2,p_3$, then at the $(N+1)^{\text{th}}$ iteration with \begin{equation}
    \mathcal{E}_g^{N+1}[\rho]=\tilde{f}\rho+\tilde{g}X\rho X+\tilde{h}Y\rho Y+\tilde{k}Z\rho Z~,
\end{equation} 
by expanding $(f+g\,x+h\,y+k\,z)(q+p_1\,x+p_2\,y+p_3\,z)$ we obtain the recursion relations
    \begin{equation}
    \begin{pmatrix}
        \tilde{f}\\ \tilde{g}\\ \tilde{h}\\ \tilde{k}
    \end{pmatrix}=\begin{pmatrix}
        q & p_1&p_2&p_3\\
        p_1&q&p_3&p_2\\
        p_2&p_3&q&p_1\\
        p_3&p_2&p_1&q
    \end{pmatrix}\begin{pmatrix}
        f\\g\\h\\k
    \end{pmatrix}:=Q\begin{pmatrix}
        f\\g\\h\\k
    \end{pmatrix}.
\end{equation}
Now $Q$ can be diagonalized as $Q=PDP^{-1}$, with
\begin{equation}
    D=\text{diag}(1,1-2p_1-2p_2,1-2p_2-2p_3,1-2p_3-2p_1).
\end{equation}
For $p_1,p_2,p_3$ that satisfy
\begin{equation}
    0<p_1+p_2<1,\quad 0<p_2+p_3<1,\quad 0<p_1+p_3<1~,
\label{eq:large-N}
\end{equation}
we have $\lim_{N\rightarrow \infty }D^N=\text{diag}(1,0,0,0)$, and  
\begin{equation}
\label{eq:diag}
    \lim_{N\rightarrow\infty} Q^N 
=P\left(\lim_{N\rightarrow\infty}D^N\right)P^{-1}=\begin{pmatrix}
        0.25 & 0.25 & 0.25 & 0.25\\
        0.25 & 0.25 & 0.25 & 0.25\\
        0.25 & 0.25 & 0.25 & 0.25\\
        0.25 & 0.25 & 0.25 & 0.25\\
    \end{pmatrix}~,
    \end{equation}
Thus for the parameters $p_1,p_2,p_3$ in  region \eqref{eq:large-N}, the coefficients $f, g, h, k$ become equal, and the large iteration of the channel flows to
\begin{equation}
\lim_{N\rightarrow \infty}    {\cal E}_g^N[\rho]=\frac{1}{4}\left(\rho+X\rho X+Y\rho Y+Z\rho Z\right)~.
\end{equation}
The iteration channel has an emergent anomalous Pauli group symmetry at large-$N$.

To better understand the requirement in \eqref{eq:large-N}, note that if the channel ${\cal E}_g(\rho)$ is generated by the composite of two elementary channels ${\cal E}_g[\rho]={\cal E}_2\circ {\cal E}_1[\rho]$ with
\begin{equation}
{\cal E}_1[\rho]=(1-c_1)\rho+c_1 X\rho X~,\quad 
{\cal E}_2[\rho]=(1-c_2)\rho+c_2 Y\rho\, Y~. 
\end{equation}
Then
\begin{equation}
    q=(1-c_1)(1-c_2),\quad  p_1= c_1 (1-c_2),\quad p_2=c_2 (1-c_1),\quad p_3= c_1 c_2.
\end{equation}
Now \eqref{eq:large-N} simply means that $0<c_1<1$ and $0<c_2<1$. This can also be intuitively understood: two frustrated (non-commuting) weak measurements injects entropy into the system.

\subsubsection{Entanglement generating channel}

For anomalous unitary symmetry $U(g)$ with $g\in G$, consider the following channel
\begin{equation}
    {\cal E}[\rho]=\frac{1}{|G|}\sum_{g\in G} U(g)\rho U(g)^{-1}~,
\end{equation}
where $|G|$ is the order of the symmetry group.
From the above discussion, the channel generates a density matrix with the anomalous average symmetries $\{U(g)\}$ for any initial density matrix $\rho$,
\begin{equation}
    U(g') {\cal E}[\rho]U(g')^{-1}=
    \frac{1}{|G|}\sum_{g\in G} U(g'g)\rho U(g'g)^{-1}={\cal E}[\rho]~,\quad \forall g'\in G~.
\end{equation}

This channel generates a density matrix whose von Neumann entropy is bounded from below by $\log d$ where $d$ is the dimension of the projective representation $U$. The dimension satisfies $d\geq d_M$ where $d_M$ is the dimension of the minimal projective representation. Thus the von Neumann entropy of the transformed density matrix satisfies
\begin{equation}
    S({\cal E}[\rho])\geq \log d\geq \log d_M~.
\end{equation}

We remark that the von Neumann entropy is bounded above by $\log d_{\cal H}$ where $d_{\cal H}$ is the dimension of the total Hilbert space ${\cal H}$. If the dimension coincides with the dimension of the minimal projective representation $d_{\cal H}=d_M$, then both the upper and lower bounds on the von Neumann entropy are saturated: $S({\cal E}[\rho])=\log d_{\cal H}$, and the channel generates the maximally mixed state. For example, in the depolarization channel discussed in \eqref{eq:decoherence} at $p=0.75$, since the von Neumann entropy is also bounded above by $\log 2$ due the dimension of the Hilbert space, and at the same time bounded from below by $\log d_M=\log 2$ due to $d_M=2$, the von Neumann entropy is exactly equal $\log 2$ and the density matrix is the maximally mixed state.

\subsection{Average symmetry preserved by quantum channel}

Let us consider initial density matrix $\rho$ with average symmetry,
\begin{equation}
    U\rho U^{-1}=\rho~.
\end{equation}
For the average symmetry to be preserved by the quantum channel, we require the Kraus operators $\{K_i\}$ of the quantum channel to satisfy
representation
\begin{equation}\label{eqn:symmetryKraus2}
    UK_jU^{-1}= \sum_{l} u_{jl}K_{l}~,
\end{equation}
where the matrix $(u_{jl})$ is a constant unitary matrix multiplied the identity operator.
In a basis of Kraus operators where the unitary matrix $(u_{ij})$ is diagonalized, it can be written as
\begin{equation}\label{eqn:symmetryKrausp2}
    UK_jU^{-1}= e^{i\phi_j }K_{j}~,
\end{equation}
where $e^{i\phi_j }$ is a phase. 
Under such a condition, the transformed density matrix also has the average symmetry:
\begin{equation}
    U{\cal E}[\rho]U^{-1}=\sum_i (UK_iU^{-1}) U\rho U^{-1}(UK_iU^{-1})^\dag=\sum_i (UK_iU^{-1}) \rho (UK_iU^{-1})^\dag
    ={\cal E}[\rho]~,
\end{equation}
where the second equality used the average symmetry of initial density matrix $U\rho U^{-1}=\rho$.

When a quantum channel preserves an average symmetry, the iterative quantum channel also preserves the average symmetry. 

As a simple example, consider the bit flip channel for systems of qubits:
\begin{equation}
    {\cal E}[\rho]=(1-p)\rho+pX\rho X~,
\end{equation}
where $X$ is the Pauli $X$ operator, $p$ is a real number between 0 and 1.
If the initial density matrix $\rho$ has average symmetry generated by Pauli $Y,Z$ operators at some site, the average symmetry is preserved by the bit flip channel. 
Since the symmetry generated by $Y,Z$ operators is anomalous, with $d_M=2$,  the transformed density matrix has von Neumann entropy bounded from below by $\log 2$.

Similarly, we can repeat the discussion with the phase flip channel with ${\cal E}[\rho]=(1-p)\rho+pZ\rho Z$ for any real number $p$ between $0$ and $1$.

\section{Application: Dynamics of Open Quantum Systems}
\label{sec:openquantumsystem}

Open quantum systems are subject to pure decoherence (dephasing) and dissipation from the environment. 
Under suitable assumptions such as Markovian process,
the time evolution of the mixed-state density matrix can be described by 
the Lindbladian master equation \cite{Lindblad:1975ef,Gorini:1975nb}: 
\begin{equation}\label{eqn:Leq}
    \frac{d\rho}{dt}={\cal L}[\rho],\quad {\cal L}[\rho]:=-i[H,\rho]+\sum_i \gamma_i\left(L_i\rho L_i^\dag-\frac{1}{2}(L_i^\dag L_i\rho+\rho L_i^\dag L_i)\right)~,
\end{equation}
where $\gamma_i\geq 0$ are constant damping rate, $L_i$ are called ``jump operators" that describe the dissipation to the environment, and $i$ denotes different types of dissipation effects. ${\cal L}[\rho]$ is called the Lindbladian acting on the density matrix $\rho$. 
We note that the Lindbladian is a linear map acting on the  density matrices.

\subsection{Symmetry of Lindbladian}

We will consider the situation when the Lindbladian ${\cal L}$ has a symmetry $G$ with (projective) representations given by the symmetry generators $\{U_a\}$.
The Lindbladian then satisfies the constraint 
\begin{equation}\label{eqn:Lindbladiansymmetry}
U_a^{-1}\circ {\cal L}\circ U_a={\cal L}~.
\end{equation}
In terms of the density matrix, this means that $U_a^{-1} {\cal L}[U_a\rho]={\cal L}[{\rho}]$. 
The projective representation $\{U_a\}$ violates the commutation relations of $G$ by nontrivial phases, and they describe a linear representation of an extension $\tilde G$.

To describe the linear relation (\ref{eqn:Lindbladiansymmetry}) more explicitly, we define the basis of the space of density matrices as $\{e_i\}$. A general density matrix can then be expressed as $\rho=\sum \rho_i e_i$, where $e_i$ represents a matrix and $\rho_i$ is a number. The action of the symmetry $U$ (where we omit the indices for different generators for simplicity) on the basis can be expressed as 
\begin{equation}
    U e_i= \sum_j U_{ij} e_j~.
\end{equation}
Since the Lindbladian operator is a linear map, we can also write 
\begin{equation}
    {\cal L}[e_i]=\sum_j {\cal L}_{ij}e_j~.
\end{equation}
The symmetry condition $U{\cal L}[\rho]={\cal L}[U\rho]$ is consequently equivalent to
\begin{equation}
    \sum_j {\cal L}_{ij}U_{jk}=U_{ij}{\cal L}_{jk}~.
\end{equation}
The above condition can be written in terms of the matrices $\mathbf{L}=({\cal L}_{ij})$, $\mathbf{U}=(U_{ij})$:
\begin{equation}
    \mathbf{LU}=\mathbf{UL}~.
\end{equation}
Thus we can apply Schur's lemma (described in sections \ref{sec:intro} and \ref{sec:Schur}), 
and it implies the spectrum of the Lindbladian has degeneracy according to the dimensions of the irreducible representations of $\tilde G$.

\subsection{Constraints on Linbladian evolution}

Suppose $\tilde G$ has
an $n$-dimensional irreducible representation with $n>1$. We can write ${\cal L}=\lambda\mathbbm{1}_n+{\cal L}'$ for some constant $\lambda$ by conjugating the density matrix with a unitary transformation, and focus on the $n$-dimensional subspace where the Lindbladian acts as $\lambda\mathbbm{1}_n$.
Sine the density matrix $\rho$ and $d\rho/dt$ are both hermitian, the constant $\lambda$ must be real. 
The density matrix restricted to the subspace satisfies
\begin{equation}
   \frac{d\rho}{dt}=\lambda \rho~.
\end{equation}
Integrating the equation, we find that the density matrix evolves as
\begin{equation}
    \rho(t)=\rho_0 e^{\lambda(t-t_0)}~,
\end{equation}
where $\rho_0=\rho(t_0)$ is the initial density matrix at time $t_0$.

We can alternatively describe the evolution using the basis $\{e_i\}$ for the density matrix introduced above, where the components of the density matrix $\{\rho_i\}$ form a vector. The component in the subspace spanned by the eigenvector with eigenvalue $\lambda$ evolves as
\begin{equation}
\frac{d\rho_i|}{dt}=\lambda\rho_i|,\quad \rho_i(t)=(\rho_0)_ie^{\lambda(t-t_0)}~. 
\end{equation}

Below we will assume the representation of the symmetry is completely reducible, such that the Lindbladian is a direct sum of blocks $\{I\}$ proportional to identity.  Then the evolution of the density matrix will be $\rho^{(I)}=\rho_0^{(I)}e^{\lambda_I t}$ for block $I$.
If $\lambda_I<0$, the corresponding blocks will evolve to 0, while the blocks with $\lambda_I>0$ will grow exponentially. On the other hand the blocks with $\lambda_I=0$ corresponds to steady states within such block.

Additionally, the evolution operator should preserve the trace of density matrix,
\begin{equation}
    1=\sum_I\text{Tr}\rho^{(I)}(t)=\sum_I e^{\lambda_I(t-t_0)}\left(\text{Tr}\rho_0^{(I)}\right)~.
\end{equation}
Let us order the spectrum of $\lambda_I$ by their distinct values $\lambda^{(1)}>\lambda^{(2)}>\cdots$, where there can be multiple blocks that share the same eigenvalue.
Then for the trace of the full density matrix to be constant in time, the trace of the density matrix in the blocks with the same nonzero $\lambda_I$ must sum to zero:
\begin{equation}\label{eqn:sumrule}
    \sum_{I:\lambda_I=\lambda\neq 0}\text{Tr}\rho_0^{(I)}=0~,
\end{equation}
for all distinct nonzero $\lambda$ in the spectrum.

Suppose the density matrix transforms in a single $n$-dimensional irreducible representation, and the symmetry constrains it to be zero ({\it i.e.} a block with $\lambda_I=0$), the density matrix describes a steady state:
\begin{equation}
    \rho(t)=\rho_0~.
\end{equation}
In such case, the symmetry implies the density matrix is a steady state of the Lindbladian evolution.

\subsubsection{Example: Lindbladian evolution of two-qubit system}

Let us illustrate the discussion with a simple example given by a system of two free qubits with a vanishing Hamiltonian. We consider the interaction with environment given by the jump operator
\begin{equation}\label{eqn:jumpexample}
    L=Z_1Z_2~,
\end{equation}
where $Z_1,Z_2$ are the Pauli $Z$ operators acting on the first and second qubit, respectively.
We will solve the simple system explicitly, and then verify the evolution satisfies the constraint (\ref{eqn:sumrule}) imposed by symmetries.

Let us parameterize a general density matrix using the computational basis ({\it i.e.} eigenbasis of the Pauli $Z$ operators)
\be\label{eqn:examplegeneralrho}
\rho = \sum_{i,j,k,l=0}^1 c_{ijkl} |ij\rangle~ \langle kl|,\quad  c_{ijkl}\in \mathbb{C}~. 
\ee
where $i,j,k,l=0,1$ mod 2. We will focus on the density matrices where $c_{ijkl}=f(i+j+k+l)$ only depends on $i+j+k+l$ mod 2. The mixed state evolves under the Lindbladian equation (\ref{eqn:Leq}) with the jump operator (\ref{eqn:jumpexample}) as
\be\label{eqn:examplelindbladian}
\frac{d\rho}{dt}={\cal L}[\rho]=
\gamma \left[\sum_{i,j,k,l} c_{ijkl} \left((-1)^{i+j+k+l}-1\right)|ij\rangle~\langle kl|\right]~.
\ee
The factor $\left((-1)^{i+j+k+l}-1\right)$ is either $0$ or $(-2)$ depending on whether $i+j+k+l$ is even or odd. Therefore the steady states are linear combinations of the states with even $i+j+k+l$.
On the other hand, the density matrix given by linear combination of the states with odd $i+j+k+l$ decreases exponentially.

Let us verify the system satisfies the constraint (\ref{eqn:sumrule}).
First, we will show the Lindbladian evolution has symmetry (\ref{eqn:Lindbladiansymmetry}) generated by all two-qubit gates that commute with $Z_1Z_2$ up to a phase.
To see this, we note that the Lindbladian evolution is
\begin{equation}
    {\cal L}[\rho]=Z_1Z_2 \rho Z_2Z_1-\rho~.
\end{equation}
The symmetry of Lindbladian evolution satisfies
\begin{equation}
    U{\cal L}[U^{-1}\rho]={\cal L}[\rho]\quad \Rightarrow \quad
    UZ_1Z_2 U^{-1}\rho Z_1Z_2-U^{-1}\rho U=Z_1Z_2 \rho Z_1Z_2-\rho~.
\end{equation}
Focusing on the density matrices where $c_{ijkl}=f(i+j+k+l)$ only depends on $i+j+k+l$ mod 2, $\rho$ has average symmetry generators $X_1, X_2, Z_1, Z_2$. Moreover, these generators commute with $Z_1Z_2$, and thus they are also symmetries of the Lindbladian. The minimal projective representation that describe the phases in the commutation relations $X_1Z_1=-Z_1X_1, X_2Z_2=-Z_2X_2$ has dimension $d_M=4$, implying that the spectrum of the Lindbladian has degeneracy of at least 4.

In this simple example, we can solve the spectrum explicitly.
The Lindbladian decompose into two $8\times 8$ blocks according to whether $i+j+k+l$ is even or odd.
Different blocks transform in different representations of the symmetry. 
In the even blocks, $\lambda=0$, and the eigenvectors are the steady states. In the odd blocks, $\lambda=-2\neq 0$, and the eigenvectors are off-diagonal since $i+j+k+l$ is odd, Thus all the blocks satisfy the sum rule (\ref{eqn:sumrule}).

\subsection{Non-dissipating states from average symmetry}

In the evolution equation (\ref{eqn:Leq}), the jump operator can be unitary. One can consider a more general situation where the jump operator is only on average unitary:
\begin{equation}
    \langle L_i^\dag L_i\rangle=\text{Tr}\left(\rho L_i^\dag L_i\right)=1~.
\end{equation}
We will consider instead a stronger operator equation, but still more general than $L_i$ being unitary:
\begin{equation}
    \left(L_i^\dag L_i-1\right)\rho+\rho \left(L_i^\dag L_i-1\right)=0~.
\end{equation}
If we take trace on both sides, this implies that $\langle L_i^\dag L_i\rangle=1$, namely the jump operator is on average unitary.
We can also regard $Q'=(L_i^\dag L_i-1)$ as an average symmetry that satisfies $Q'\rho=-\rho Q'$, but $Q'$ is not necessarily invertible (when $L_i$ is exactly unitary, $Q'=0$).

In such case, we can rewrite the equation as follows: define $\rho=e^{-iHt}\tilde\rho e^{iHt}$, the equation reduces to
\begin{equation}\label{eqn:openquantumevo}
    e^{-iHt}\frac{d\tilde \rho}{dt}e^{iHt}=\sum_i \gamma_i\left(L_i \rho L_i^\dag-\rho\right)~.
\end{equation}

Moreover, let us focus on the case where $L_i$ is an average symmetry. Thus the conditions on $L_i$ are
\begin{equation}
\text{Non-dissipating}:\quad     L_i^\dag \rho L_i=\rho,\quad  \left(L_i^\dag L_i-1\right)\rho=-\rho\left(L_i^\dag L_i-1\right)~.
\end{equation} 
These conditions imply that the right hand side of (\ref{eqn:openquantumevo}) vanishes, $\frac{d\tilde\rho}{dt}=0$.
Thus we find non-dissipating state $\rho$ that evolves as in a closed unitary system from the time-independent steady state $\tilde\rho_0$ according to the Hamiltonian $H$
\begin{equation}
    \rho=e^{-iHt}\tilde\rho e^{iHt}~.
\end{equation}

\section{Conclusion and Outlook}
\label{sec:discussion}

In this work we consider the mixed anomaly of average/weak symmetries and their effects on  mixed-state density matrices as well as their evolutions. In particular, when the generators of the average symmetry do not commute with each other, a density matrix that satisfy these symmetries can host degeneracies in the spectrum. We consider examples where the symmetries of the XY spin chain, Heisenberg chain and the toric code models are imposed. We also discuss the implications of average symmetries on dynamics due to quantum channels and Lindbladian evolution. Below we will comment on the subtleties, relation with other works as well as future directions. 

\begin{itemize}

\item A common approach for studying density matrix is using the ``doubled'' Hilbert space, by converting the density matrix $\rho=\sum_{i,j}\rho_{ij}|i\rangle \langle j|$ into the pure state $\sum_{ij}\rho_{ij}|i\rangle\otimes |j\rangle$ in the doubled Hilbert space ${\cal H}\otimes {\cal H}^*$ for the original Hilbert space ${\cal H}$ \cite{CHOI1975285}. However, the anomalous symmetry as described by a projective representation in ${\cal H}$ becomes a linear representation acting on the doubled Hilbert space ${\cal H}\otimes {\cal H}^*$. Thus the anomaly discussed here is not manifest in the doubled Hilbert space formalism. It would be interesting to see how the anomaly appears in the doubled Hilbert space formalism.

\item The discussion focused on mixed states in lattice system with finite dimensional Hilbert space, but it would be interesting to explore the implications in continnum quantum field theory, such as anomalous average symmetry in conformal field theory and the holography correspondence. 
\item We briefly comment on the connections between our discussions of steady states for open systems in Section \ref{sec:openquantumsystem} versus other similarly flavored results. Quantum many-body scars, originally discovered in \cite{bernien2017probing,PhysRevB.98.235155} (see \cite{2021NatPh..17..675S,moudgalya2022quantum} for reviews), correspond to a discrete number of non-ETH eigenstates that scar a seemingly ergodic system. Therefore, if special initial states were chosen, their time evolution would be very different from usual thermal states. These features are usually linked to the existence of dynamical symmetries \cite{PhysRevLett.125.230602,PhysRevB.102.041117,PhysRevResearch.2.043305,PhysRevLett.126.120604} and Hilbert space fragmentation \cite{PhysRevX.10.011047,PhysRevB.101.174204, Moudgalya_2021,10.21468/SciPostPhysCore.4.2.010}. We would like to remark that the quantum many-body scarred states are eigenstates of the Hamiltonian, and the dynamical symmetries are emergent symmetries not present in the Hamiltonian. In comparison, our setup focused directly on the density matrices which are not necessarily eigenstates of a certain Hamiltonian. However, similar to the discussions of quantum scars, our setup also does not directly involve symmetries of some Hamiltonian: although the definition of Lindbladian includes a Hamiltonian piece, our results relied only on the symmetry properties of the full Lindbladian.

\item We have derived a new bound on the von Neumann entropy and R{\'e}nyi entropy from anomalous average symmetry. On the other hand,  Neumann entropy captures not only quantum entanglement but also classical correlations. It would be interesting to derive new bounds on other measures of quantum entanglement such as entanglement negativity \cite{PhysRevA.58.883} (see {\it e.g.} \cite{Calabrese:2012ew,PhysRevA.88.042318,PRXQuantum.2.030313} for some applications).

\item In Section \ref{sec:toric}, we discussed the mixed anomaly between two average 1-form symmetries. 
The discussion, however, is not obviously related to mixed state topological orders. For example in the context of toric code under error channels \cite{Dennis_2002,bao2023mixedstate,fan2023diagnostics,PRXQuantum.4.030317}, the bit-flip channel exhibits the all the weak one-form symmetries discussed in our work. However, if the initial state is a ground state of the toric code Hamiltonian which has a fixed eigenvalue of two of the four symmetry generators, then neither this initial state nor the final state after the channel exhibits all the weak one-form symmetries, but only half of them.

\item It would interesting to compare our bound with the bound on topological entanglement entropy in \cite{Kim_2023}. Since topological orders generate an anomalous symmetry (which may not be invertible), where the anomaly comes from the remote detectability {\it i.e.} the topological operators have nontrivial braiding statistics,
the bound on the topological entanglement entropy in terms of the total quantum dimension of the topological order \cite{Kim_2023} is expected to come from a similar reasoning.

\item In unitary continuum relativistic quantum field theories the renormalization group flows obey monotonic theorems such as $c$-theorem, $F$-theorem and $a$-theorem, where a function is monotonic along the renormalization group flow.
At the fixed point of the renormalization group flow, the function coincides with suitable entanglement entropy of the theory \cite{Casini_2004,Casini_2007,Casini_2012,solodukhin2013atheorem,Nishioka_2018}. On the other hand, there can be emergent symmetry at the end of the renormalization group flows. It would interesting to use our method to explore the constraints from anomalous average symmetry on the renormalization group flow.

\item Although the discussions focus on symmetries that have an inverse transformation, many quantum systems on lattice have non-invertible symmetries, including (1+1)d critical Ising model \cite{Aasen_2016,Seiberg:2023cdc}, (1+1)d anyon chain models \cite{Feiguin_2007,Trebst_2008}, and lattice gauge theories in (3+1)d \cite{Koide:2021zxj,Choi:2021kmx,Barkeshli:2022edm,Cordova:2023bja}.
It is curious to understand similar constraints for non-invertible symmetries such as Kramers--Wannier duality, by investigating the analogue of Schur's lemma for non-invertible symmetry.

\item Symmetries in pure state systems can be described from a bulk topological order (see {\it e.g.} \cite{Kong_2020,kaidi2023symmetry,freed2023topological}). In particular, bulk topological order plays an important role in constraining the dynamics of quantum system using the corresponding symmetry \cite{Zhang:2023wlu,Cordova:2023bja}.
For instance, anomalous symmetries can be described by nontrivial SPT phase in the bulk. It would be interesting to establish the  connection between average symmetry in mixed states on the boundary and corresponding bulk properties.

\item It would be interesting to generalize the argument to constrain whether the system is gapless. For instance, we can use density matrix of the ground states and arguments similar to \cite{PhysRevLett.84.3370}. In one spatial dimension or higher, we need to introduce symmetry fluxes to detect anomalies using commutation relations of the symmetry generators. Following \cite{PhysRevLett.84.3370}, we can (1) assume there is gap, (2) then insert symmetry flux with ``winding'' that adiabatic deforms the states from the original Hilbert space to the twisted Hilbert space and back to the original Hilbert space, and (3) exam if the symmetry generators with fluxes obey nontrivial commutation relations.
Using such an approach, we can conclude either there is a degeneracy in the density matrix, or the system does not have a gap, or the average symmetry is absent.

\item While composing the paper, reference \cite{zhou2023reviving} appeared, showing that when coupling a clean system with Lieb--Schultz--Mattis anomaly to a bath, such that the resultant is short-range correlated, the Lieb--Schultz--Mattis theorem can still be visible from the degeneracy in the spectrum of the entanglement Hamiltonian, obtained from tracing out the bath. The degeneracy in the reduced density matrix can be indicated from the anomalies between symmetries acting on the reduced density matrix in the same fashion as discussed in this work.

\end{itemize}

\section*{Acknowledgements}

We thank 
Zhen Bi, Meng Cheng, Ceren Dag, Tyler Ellison,
Tarun Grover, Tim Hsieh, Luca Iliesiu, Andreas Karch, Itamar Kimchi, Zhi Li, Alexey Milekhin, Subir Sachdev, Henry Schackleton, Julian Sonner, Alex Turzillo, Ruben Verresen, Ashvin Vishwanath, Chong Wang, Yifan Wang, Cenke Xu, Zhenbin Yang, Jian-Hao Zhang for discussions.
We thank 
Andrea Antinucci, Meng Cheng, Giovanni Galati, Alexey Milekhin, Shu-Heng Shao, Ryan Thorngren, Alex Turzillo, Cenke Xu and Carolyn Zhang for comments on a draft.
P.-S.H. is supported by Simons Collaboration on Global Categorical Symmetries. Z.-X.L. is supported
by the Simons Collaboration on Ultra-Quantum Matter with Grant 651440 from the Simons Foundation.
H.-Y.S. is supported in part by the U.S. Department of Energy under Grant No. DESC0022021, and Grant 651440 from the Simons Foundation. 

\appendix

\section{Schur's Lemma and Corollaries}
\label{sec:Schur}

Below we will elaborate the mathematical observation summarized in Section \ref{sec:intro}. We will start with two special cases and then move on to the general situation. 

Consider an $N$-by-$N$ matrix $M$, a finite group $G$ generated by $\{g_a\}$, and an $N$-dimensional representation $U(g_a)$ of these generators. Assume that $M$ satisfies 
\be
U(g_a) M = M U(g_a),\quad \forall a. 
\label{eq:main}
\ee
We now derive the implications on $M$ based on the properties of $U$. 
\begin{itemize}
\item[(i)]
If the representation $U$ is irreducible, then by Schur's Lemma, $M$ must be a constant multiple of the identity matrix. Consequently, $M$ has only one distinct eigenvalue with degeneracy $N$. 

\item [(ii)] Next we consider the case where $U$ is equivalent to a semi-simple representation, \textit{i.e.}, when written in a suitable basis, it is a direct sum of irreducible representations. In other words, there exists $V$ such that
\be
U'(g_a):= V^{-1} U(g_a) V = \oplus_i U_i(g_a),\quad \forall a, 
\ee
where $U_i$ is irreducible with dimension $n_i$.  Such $U$ is also called completely reducible. Equation \eqref{eq:main} can then be equivalently written as: 
\be
U'(g_a)M'=M' U'(g_a),\quad M'=  V^{-1} M V. 
\label{eq:similarity}
\ee
\begin{itemize}
    \item[$\bullet$] To gain some intuition on Equation \eqref{eq:similarity}, we start from the simplest two-block example:
    \be
    U'(g_a)=\left( \begin{matrix} {U}_1(g_a) & 0 \\ 0 & {U}_2(g_a) \end{matrix}\right),\quad M'=\left( \begin{matrix} A & B \\ C & D \end{matrix}\right),
    \ee
    where $A$ has the same dimensions as ${U}_1(g_a)$ and $D$ has the same dimensions as ${U}_2(g_a)$.
    Equation \eqref{eq:similarity} then requires
    \be
    \left( \begin{matrix} {U}_1(g_a) A &{U}_1(g_a) B \\ {U}_2(g_a) C &\ {U}_2(g_a) D \end{matrix}\right)=
    \left( \begin{matrix} A {U}_1(g_a)  &\ B {U}_2(g_a)  \\ C {U}_1(g_a)  &\ D {U}_2(g_a)  \end{matrix}\right),\quad \forall a.
    \label{eq:matrix_eqn}
    \ee
    It immediately follows that the two diagonal blocks $A$, $D$ must be proportional to identity by Schur's lemma. As for the off-diagonal blocks, the lemma further indicates that (1) if $n_1\neq n_2$, then $B=C=0$. Then our original matrix of interest, $M$,  is equivalent to the direct sum of two diagonal matrices, each proportional to identity; (2) while if $n_1=n_2$, another possibility is allowed: ${U}_1$ and ${U}_2$ are equivalent and related by the similarity transformation given by the invertible matrix $B=C^{-1}$. In this case, 
    \be
    {M}'= \left( \begin{matrix} \lambda_1\mathbbm{1}  &\ B  \\ B^{-1}  &\ \lambda_2\mathbbm{1}  \end{matrix}\right),\  \det[{M}'-\lambda \mathbbm{1}]=(\lambda_1-\lambda)^{n_1} (\lambda_2-\lambda-\lambda_1^{-1})^{n_2},
    \label{eq:det1}
    \ee
    where we have used the general relation for block matrix
    \be
    \det \left( \begin{matrix} A  &\ B  \\ C  &\ D \end{matrix}\right)=\det (A) \det (D-CA^{-1}B)
    \label{eq:det_id}
    \ee
    when $A$ is inversible. Equation \eqref{eq:det1} indicates that ${M}'$ has eigenvalue $\lambda_1$ with degeneracy $n_1$ and eigenvalue $\lambda_2-\lambda-\lambda_1^{-1}$ with degeneracy $n_2$. Put it in another way, $M$ is equivalent to a matrix which is a direct sum of two submatrices, each proportional to identity. 
    
    \item[$\bullet$] Going beyond two blocks, one can repeatedly use equation \eqref{eq:det_id} to show that $M$ is always equivalent to direct sum of matrices that are proportional to identity, and the dimensions of the identity matrices follow the dimensions of the ${U}_i$ blocks. We leave the proof as an exercise for the readers. 
\end{itemize}

\item [(iii)] Next we move on to the case where $U$, in a certain basis $U'$, is decomposable into one irreducible representation ${U}_1$ of $G$, and some reducible representation ${U}_2$ of $G.$ Again we write ${M}'$ in block form and solve \eqref{eq:matrix_eqn}. Since ${U}_1$ is irreducible, $A$ must be proportional to identity. Generically we have $n_1\neq n_2$, so $B$ and $C$ are not invertible. However, we can still derive that ${U}_1 B C= B {U}_2 C=BC {U}_1$, \textit{i.e.}, $BC$ intertwines with the irreducible representation ${U}_1$, so $BC$ must be proportional to identity. Since $A$ is invertible, using again the identity \eqref{eq:det_id}, we can derive that in a certain basis, $M$ has an isolated identity block of dimension $n_1$. 

\end{itemize}

\bibliographystyle{utphys} 
\bibliography{biblio}

\end{document}